**SentRNA: Improving computational RNA design by incorporating a prior of human design strategies**

Jade Shi[1], EteRNA players, Rhiju Das[2], and Vijay S. Pande[3*]
1. Department of Chemistry, Stanford University, jadeshi@stanford.edu
2. Department of Biochemistry, Stanford University, rhiju@stanford.edu
3. Department of Bioengineering, Stanford University, pande@stanford.edu

**Abstract:**
Solving the RNA inverse folding problem is a critical prerequisite to RNA design, an emerging field in bioengineering with a broad range of applications from reaction catalysis to cancer therapy. Although significant progress has been made in developing machine-based inverse RNA folding algorithms, current approaches still have difficulty designing sequences for large or complex targets. On the other hand, human players of the online RNA design game EteRNA have consistently shown superior performance in this regard, being able to readily design sequences for targets that are challenging for machine algorithms. Here we present a novel approach to the RNA design problem, SentRNA, a design agent consisting of a fully-connected neural network trained end-to-end using human-designed RNA sequences. We show that through this approach, SentRNA can solve complex targets previously unsolvable by any machine-based approach and achieve state-of-the-art performance on two separate challenging test sets. Our results demonstrate that incorporating human design strategies into a design algorithm can significantly boost machine performance and suggests a new paradigm for machine-based RNA design.

**Introduction:**
Designing RNA molecules to perform specific functions is an emerging field of modern bioengineering research[1,2,3,4,5] with diverse biological applications from cancer therapy[1] to intracellular reaction catalysis.[3] Because an RNA molecule's function is highly dependent on its structure, in order to effectively design RNA molecules to perform specific functions, one must first solve the RNA inverse folding problem: given a target structure, predict an RNA sequence that will fold into that structure. As such, significant efforts have been made over the past several decades in developing computational algorithms to reliably predict RNA sequences that fold into a given target.[6,7,8,9,10,11,12,13,14,15,16]

A large fraction of currently available inverse RNA folding algorithms follows the general pipeline of first generating an initial guess of an RNA sequence and the refining that sequence using some form of stochastic search. Algorithms that fall under this category include RNAInverse,[6] RNA-SSD,[7] INFO-RNA,[8] NUPACK,[10] MODENA,[11] Frnakenstein,[13] and ERD.[14] RNAInverse, one of the first inverse folding algorithms, initializes the sequence randomly and then uses a simple adaptive walk to randomly sample better sequences. RNA-SSD first performs hierarchical decomposition of the target and then performs adaptive walk separately on each substructure to reduce the size of the search space. INFO-RNA first generates an initial guess of the sequence using dynamic programming to estimate the minimum energy sequence for a target structure, and then performs simulated annealing on the sequence. NUPACK performs hierarchical decomposition of the target and assigns an initial sequence to each substructure. For each sequence, it then generates a thermodynamic ensemble of possible structures and stochastically perturbs the sequence to optimize the "ensemble defect" term, which represents the average number of improperly paired bases relative to the target over the entire ensemble. MODENA and Frnakenstein first generate an ensemble of candidate sequences and then use genetic algorithms to optimize complex objective functions. Finally, ERD generates an initial sequence by decomposing the target structure into substructures, and then randomly assigns a naturally occurring subsequence drawn from a public database to each substructure. Defects in the full target structure given this intialization are then corrected using an evolutionary algorithm.

On the other hand, there are also several algorithms that do not follow this formula. For instance, DSS-Opt[9] foregoes stochastic search and instead attempts to generate a valid sequence directly from gradient-based optimization of an objective function that includes the predicted free energy of the target and a "negative design" term that punishes improperly paired bases. antaRNA[16] employs "ant-colony" optimization, in which a sequence is first generated via a weighted random search, and the goodness of these sequences is then used to refine the weights and improve subsequent sequence generations. IncaRNAtion[12] first generates a GC-weighted partition function for the target structure, and then adaptively samples sequences from it to match a desired GC content. Finally, RNAiFold[15] employs constraint programming that exhaustively searches over all possible sequences compatible with a given target.

However, despite the significant progress in developing inverse RNA algorithms, current algorithms consistently have difficulty designing sequences for particularly large or complex targets.[17] On the other hand, human playeyrs of the online RNA design game EteRNA[18] have shown consistently superior performance to machine-based algorithms for such targets. Players of the game are shown 2D representations of target RNA structures ("puzzles") and asked to propose sequences that fold into them. These sequences are first judged using the ViennaRNA 1.8.5 software package[6] and then validated experimentally. Through this cycle of design and evaluation, players build a collective library of design strategies through visual pattern recognition that can then be applied to new, more complex puzzles. Remarkably, these human-developed strategies have proven very effective for RNA design. For example, EteRNA players significantly outperform computational algorithms on the Eterna100, a set of 100 challenging puzzles designed by EteRNA players to showcase a variety of RNA structural elements that make design difficult.[17] A recent benchmark against 6 different inverse folding algorithms showed that while top-ranking human players can solve all 100 puzzles, even the best-scoring computational algorithm, MODENA, could only solve 54 / 100 puzzles. Given the success of these human strategies, we investigate whether incorporating these strategies into a design algorithm can improve machine performance past the current state of the art.

We present SentRNA, a computational agent for RNA design that significantly outperforms existing methods by learning human-like design strategies in an end-to-end, data driven manner. The agent consists of a fully-connected neural network that takes as input a featurized representation of the local environment around a given position in a puzzle. The output is length-4, corresponding to the four RNA nucleotides (bases): A, U, C, or G. The model is trained using the *eternasolves* dataset, a custom-compiled collection of $1.8 \times 10^4$ player-submitted solutions across 724 unique puzzles. These puzzles comprise both the "Progression" puzzles, designed for beginning EteRNA players, as well as several "Lab" puzzles for which solutions were experimentally synthesized and tested. During validation and testing the agent takes an initially blank puzzle and assigns bases to every position greedily based on the output values. If this initial prediction is not valid, as judged by ViennaRNA 1.8.5, it is further refined via an adaptive walk using a canon of standard design moves compiled by players and taught to new players through the game's tutorials.

Overall, we trained and tested an ensemble of 154 models, each using a distinct training set and model input (see Methods). Collectively, the ensemble of models can solve 47 / 100 puzzles from the Eterna100 by neural network prediction alone, and 78 / 100 puzzles using neural network prediction + refinement. To address the possibility of our method overfitting to EteRNA-like puzzles, we also tested SentRNA on an independent set of 63 targets recently used by Garcia-Martin et al. to benchmark a set of 10 inverse design algorithms that comprise structures taken from the Rfam 9.0 and GenBank database, as well as an additional set of longer, naturally-occurring biological RNA structures.[19] Despite being a test set independent from EteRNA, we find that SentRNA is able to achieve state-of-the-art performance here as well. This study demonstrates that incorporating human design strategies into a computational RNA design agent can lead to significant increases in performance over previous methods, and represents a new paradigm in machine-based RNA design.

**Methods:**
Code availability:
The source code for SentRNA, all our trained models, and the full *eternasolves* dataset can be found on GitHub: https://github.com/jadeshi/SentRNA.

Hardware:
We performed all model training and validation using a desktop computer with an Intel Core i7-6700K @ 4.00 GHz CPU and 16 GB of RAM. Initial testing and refinement on the Eterna100 using the full refinement moveset was done using this same machine. Subsequent refinement using restricted refinement movesets (see Results), as well as testing and refinement of the 63 non-EteRNA targets from Garcia-Martin et al.[19] was run on the Sherlock 2.0 computing cluster at Stanford University,[20] with each instance of model refinement utilizing a Intel Xeon E5-2640v4 processor and 1 GB of RAM.

Creating 2D structural representations of puzzles:
During training and testing, we used a custom rendering method, which we hereafter call EteRNA rendering, to translate puzzles to 2D structures given their dot-bracket representations. This rendering method reproduces exactly what human players see in-game when solving the structure in EteRNA. We believe this is the most natural representation method for this study since we are training SentRNA on data submitted by these human players.

Assessment of RNA sequences:
We assess whether a given RNA sequence folds into the target structure using Vienna 1.8.5. We chose this version over newer versions such as Vienna 2 for consistency, because Vienna 1.8.5 is the version currently implemented in EteRNA, and is therefore the natural choice to assess whether SentRNA is capable of learning and generalizing human design strategies from EteRNA player solutions.

Neural network architecture:

Our goal is to create an RNA design agent that can propose a sequence of RNA bases that folds into a given target structure, i.e. fill in an initially blank EteRNA puzzle. To do this, we employ a fully connected neural network that assigns an identity of A, U, C, or G to each position in the puzzle given a featurized representation of its local environment. During test time, we expose the agent to every position in the puzzle sequentially and have it predict its identity. The neural network was implemented using TensorFlow[21] and contains three hidden layers of 100 nodes with ReLU nonlinearities. The output is length-4, corresponding to the four RNA bases: A, U, C, and G. During validation and test time, base identities are assigned greedily to the puzzle based on these output values.

Given a position $x$ in the puzzle, the input for this position to the agent is a combination of information about its bonding partner, nearest neighbors, and long-range features, which can include, for example, next nearest neighbors or adjacent closing pairs in a multiloop. While the bonding partner and nearest neighbor information is provided to the agent by default, long-range features are learned through the training data.

The information about the bonding partner is encoded as a length-5 vector, with each position in the vector representing either A, U, C, G, or "none" (i.e. a blank position that does not have a base assigned to it yet). A value of 1 is assigned to the position corresponding to the identity of the bonding partner, while all other values are set to 0. If there is no bonding partner, all values are set to 0. The nearest neighbor information is encoded as a length-11 vector, a combination of two length-5 one-hot vectors corresponding to the identities of the bases directly before and after it in the sequence, and a single value that corresponds to the angle in radians formed by the base and its nearest neighbors. This angle serves to distinguish bases belonging to different substructures in the puzzle. For example, a base situated in the middle of a large internal loop will have a larger angle than a base positioned in a 4-loop. As a design choice, any position in the middle of a stack of bonded bases was assigned an angle of 0. Also, if the model is looking at either the first or last position in the puzzle, the "before" and "after" nearest neighbor portions of the input respectively are set to 0.

Long-range features refer to important positions $y$ in the puzzle relative to $x$ that the agent should also have knowledge of when deciding what base to assign to $x$. These are each defined by a set of two values: 1) the Cartesian distance, $L$, between $x$ and $y$ in the puzzle given the 2D rendering of the puzzle, and 2) the angle in radians, $\Phi$, formed by positions $x - 1$, $x$, and $y$. These two values are stored in a list, [$L$, $\Phi$], and serve as a label for the feature. For example, a label of [23.0, 1.6] corresponds to a base's bonding partner in the middle of a stem. The bonding distance is equal to 23.0 EteRNA rendering distance units, and the angle between the previous base in the stem, the current base, and the bonding partner is 1.6 radians, or 90 degrees. A length-5 vector of zeros is then appended to the input vector to serve as a placeholder for the feature. During training, validation, or testing, when the agent is looking at a given position $x$, it computes $L$ and $\Phi$ between $x$ and all other positions $y_i$ in the puzzle, and if both L and $\Phi$ match that of a long-range feature used in the model within some threshold, a 1 is assigned to the corresponding placeholder depending on the identity of $y_i$ (A, U, C, G, or "none"). We set the threshold for both $L$ and $\Phi$ to an arbitrary small value of $10^{-5}$.

We determine what long-range features to use (i.e. which features should be considered "important") using a mutual information metric over player solutions. First, we perform a pairwise mutual information calculation using all the player solutions for a given puzzle to form a $l$ x $l$ mutual information matrix, where $l$ is the length of the puzzle. We then select the top $M$ (user-defined) positions in the matrix with highest mutual information, and for each of these positions (*x, y*) compute $L$ and $\Phi$ to give a list of long-range features for the puzzle (Figure 1). This process is repeated for each puzzle, and the unique long-range features across all the puzzles are then combined into an aggregate list of long-range features. A random subset of $N$ (user-defined) features is then selected from this list and used to define a model to be used for training, validation, and testing.

      By defining long-range features using a mutual information metric, our goal is to provide additional useful prior information to the model. High mutual information between positions x and y indicates that the identity of position *x* is strongly correlated to that of position *y*. This suggests one of two possibilities: 1) when EteRNA players are choosing in-game what to assign for *x*, they are first typically looking at *y*, or vice versa due to their knowledge of the structure, or 2) the two positions are constrained to be correlated by the biophysyical and energetic constraints of the target. Therefore, by only including positions with high mutual information to the agent's field of vision, we are providing either human prior information or biophysical prior information into the agent, allowing it to prioritize what humans, or nature, have deemed to be important. As a result, we provide the model with enough information to prevent underfitting and enable it to apply its learned strategies to more difficult puzzles. On the other hand, we also limit the model complexity such that we can train the model using a relatively small number of training examples without overfitting.

      We observed that the inclusion of instructive long-range features allowed SentRNA to learn human design strategies much more easily. For example, the puzzle Shortie 6 from the Eterna100 requires a strategy called 4-loop boosting to solve, which describes the process of mutating the N-terminal end of a 4-loop to a G (Figure 1). This mutation energetically stabilizes the preceding stem and is a necessary step to solve the puzzle, and is one of the most commonly used strategies in EteRNA. We found that including a long-range feature between the N-terminal base and the opposing C-terminal base of the loop (Figure 1, right inset, blue) in SentRNA's input allows it to uniquely identify the 4-loop and and reliably learn to mutate the N-terminal 4-loop base to G; 99.7% of the 306 models which contained this long-range feature successfully learned the 4-loop boost. On the other hand, if this feature is not present, we observed that it was much more difficult for SentRNA to learn this strategy; only 2.4% of the 328 models that did not contain this feature learned the 4-loop boost. These results suggest that the nearest neighbor features (Figure 1, insets, red) are in most cases inadequate to distinguish the N-terminal 4-loop base from other locations in training puzzles in which G is generally not present, such as 3-loops or the middle of internal loops. As a result, the training on this base is contaminated by other unrelated bases, and SentRNA is lead to mistakenly believe that the N-terminal 4-loop position should be A.

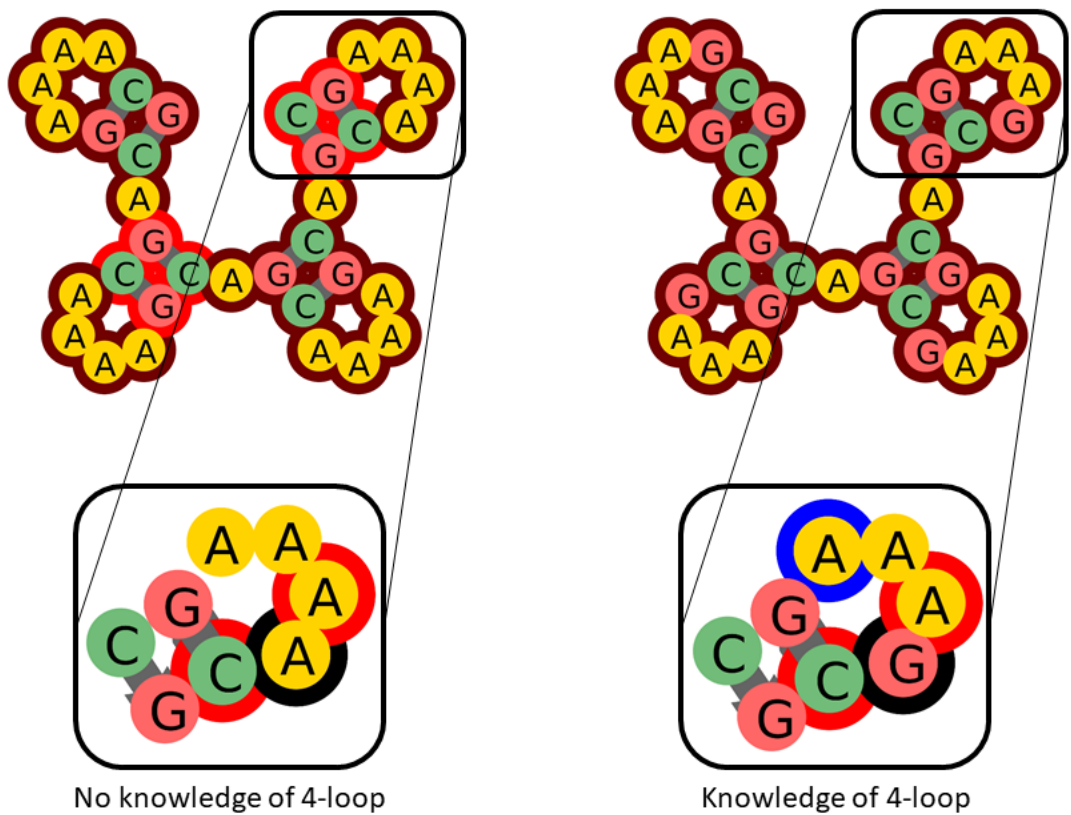

**Figure 1:** SentRNA can learn to boost 4-loops with a G at the N-terminal end much more effectively if there is a long-range feature (right inset, blue) from to the C-terminal position of the 4-loop that uniquely identifies the 4-loop. On the other hand, if this feature is not present, by relying on only its nearest neighbor features (insets, red), SentRNA cannot distinguish between this position and the other positions in the puzzle whose nearest neighbors define a similar angle (e.g. 3-loops or the middle of internal loops). This leads to contamination during training and prevents SentRNA from learning how to uniquely boost the N-terminal 4-loop position with a G. As a result, SentRNA struggles to solve puzzles such as Shortie 6 from the Eterna100 in which boosting is a necessary solution step (left).

Finally, we note that SentRNA's architecture and mechanism of action are analogous to the process of using a 2D convolutional neural network to recognize images, where convolution kernels are used to scan through the image and detect specific features in the data. Here, we are instead scanning a sparse 2D convolution kernel through the EteRNA rendering ("image") of the puzzle to recognize important structural motifs. Our kernel is sparse because only the subset of positions that show high mutual information as calculated from the training data have nonzero weights that are trainable, while all other positions are forced to have zero weights. In other words, we are using the prior information from the training data to impose hard constraints on the complexity of our kernel (Figure 2).

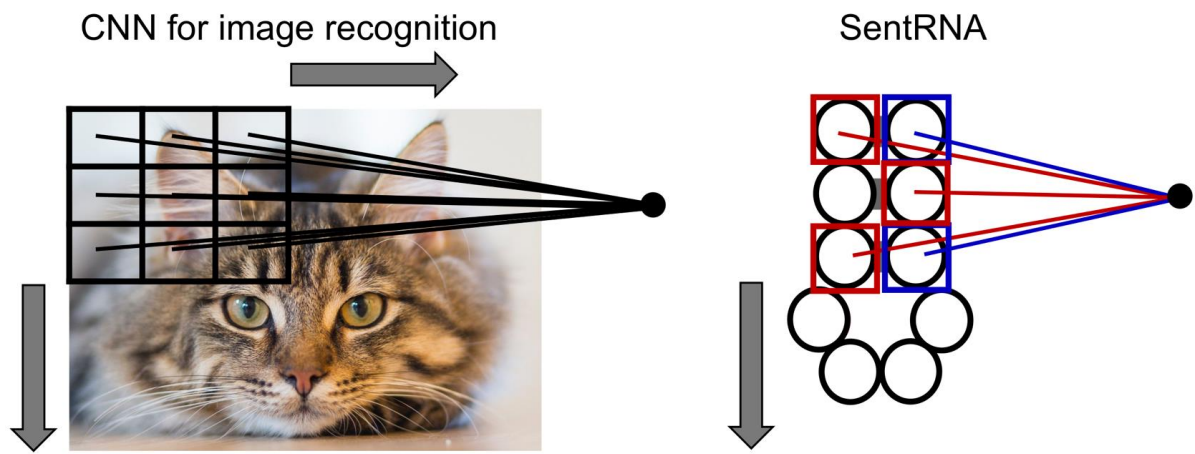

**Figure 2:** SentRNA (right) operates analogously to a convolutional neural network (CNN) for image recognition (left). In contrast to a standard CNN, however, instead of scanning a dense convolution kernel over an image, SentRNA uses a sparse kernel that detects bonding pairs and nearest neighbors (red), as well as locations of high mutual information as computed using the training data (blue). Using the kernel, SentRNA scans through the 2D representation of the RNA structure generated by EteRNA rendering and assigns bases to each position (see Figure 3 for a more detailed explanation of SentRNA's architecture and mechanism).

Training algorithm:
We used subsets of the first 721 / 724 puzzles from *eternasolves* to train the model, puzzles 722 and 724 for initial validation and testing respectively, and the Eterna100 for more extensive testing. Puzzle 723 was skipped due to being completely unstructured and not useful for validation. We confirmed that there was no contamination between the training, validation, and test sets.

Because there is no straightforward way to determine *a priori* what long-range features and training examples will result in the best-performing model, we decided to train and test an extensive ensemble of models. To do this, we first computed an aggregate list of 42 long-range features using all puzzles from *eternasolves* with at least 50 submitted solutions, allowing each puzzle to contribute only one long-range feature ($M$ = 1). We set this threshold of 50 player solutions since puzzles with a small number of solutions can introduce noise into the mutual information calculation. We then randomly selected a subset of long-range features from the aggregate list and built a model using these features. We built one model each using between 0 to 42 randomly chosen features, and repeated this process multiple times to build a total of 860 models. To form the training sets for these models, we first randomly chose 50 puzzles from *eternasolves* to serve as training puzzles. This gives us 50 lists of player solutions, one for each puzzle. We then take the first solution from each list to form a training set of 50 player solutions.

To train each model, we use the following procedure. For each player solution in our training set, we first visually render the corresponding puzzle using the EteRNA rendering method. We then set the identity of every position in the puzzle to the corresponding base in the player solution, and featurize each position into bonding pair, nearest neighbors, and long-range features to form the input vector. The output label is set to the identity of the corresponding position in the player solution. Then, we decompose the player solution into a "solution trajectory" to teach the agent how to solve an initially blank puzzle with no bases assigned (i.e. during test time). This is done by first removing all base assignments from the puzzle. A position in the puzzle is then selected and featurized. All input features (bonding partner, nearest neighbors, long-range features) are at this point set to "none", and the output label is set to the identity of the corresponding position in the player solution. This position is then filled in with the player solution base, and the next position is picked and featurized (Figure 3). This process continues sequentially until all positions in the puzzle have been featurized. This process is essentially mimicking the process of a human player filling in the puzzle sequentially base by base and training the model to reproduce these steps. During validation and testing, the agent proceeds through each position in the (initially blank) puzzle sequentially and assigns bases greedily based on the model outputs.

We note that when using this sequential fill-in approach, we are only interested in being able to reproduce the final player solution, not the steps taken by that player to reach the solution. This was an intentional choice to avoid noise in the training process. Through discussions with many EteRNA players, we found that the exact process taken by a player to solve a puzzle in-game is often much longer and more convoluted than simply filling in the puzzle base-by-base sequentially. For example, a common strategy employed by players it to putatively assign sets of bases to the puzzle, and then mutate these bases at a later stage of solving to refine the solution. Therefore, training an agent to reproduce exact solution trajectories would likely result in many unnecessary, unproductive moves that would later have to be undone, and potentially even result in infinite loops of assigning and unassigning bases in certain cases. To avoid these potential situations, we opted to use sequential fill-in as a simple and consistent (albeit artificial) means to reach the final solution.

We initialize each model using Gaussian weights ($\mu=0$, $\sigma=0.02$), unit biases, and a learning rate of 0.001. We train each model using the Adam optimizer[22] for a total of 1000 epochs, performing a validation on puzzle 722 every 100 epochs, to give a total of 10 models. The model with the highest validation accuracy is then used for testing on puzzle 724. During validation and testing, we allowed the model two attempts at predicting a sequence, once using a blank sequence as input, and again using the initial model-predicted sequence as input. The second attempt is intended as an opportunity for the model to refine its first prediction. If the model proposed valid solutions for both validation and test puzzles, it was then subjected to more extensive testing on the Eterna100.

In total, we trained 860 models, and of these models, 802 models passed initial validation and testing on puzzles 722 and 724 of *eternasolves*. Of these 802 models, we then further tested 154 of these models on the full Eterna100 and the 63 structures from Garcia-Martin et al.[19] We decided to only test a subset of our trained models on the Eterna100 due to the fact that the overall performance of the ensemble, measured by the total number of puzzles solvable across the ensemble, had firmly plateaued at a stable value of 78 / 100 by this point. We therefore concluded that additional testing of models would likely be unproductive in terms of enhancing performance.

1. Select puzzle

2. Compute pairwise mutual information using player solutions and note large values

3. Define long-range features

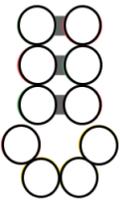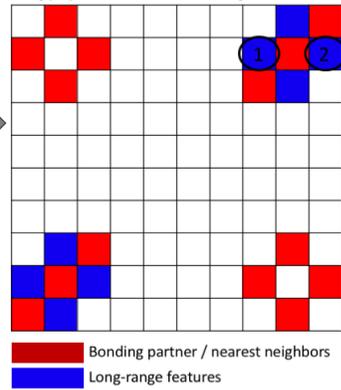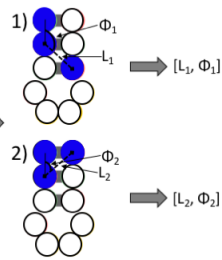

Bonding partner / nearest neighbors
Long-range features

4. Train with full player solution and solution trajectory

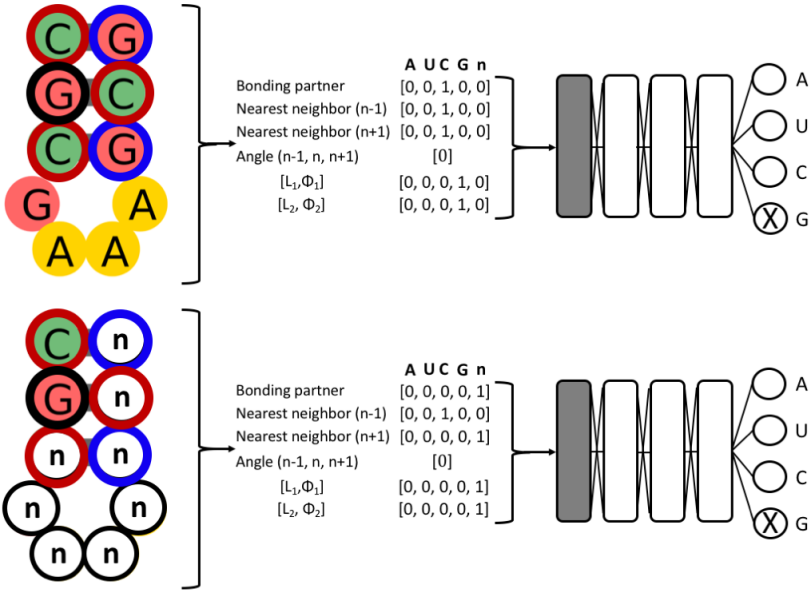

5. Validate on new puzzle

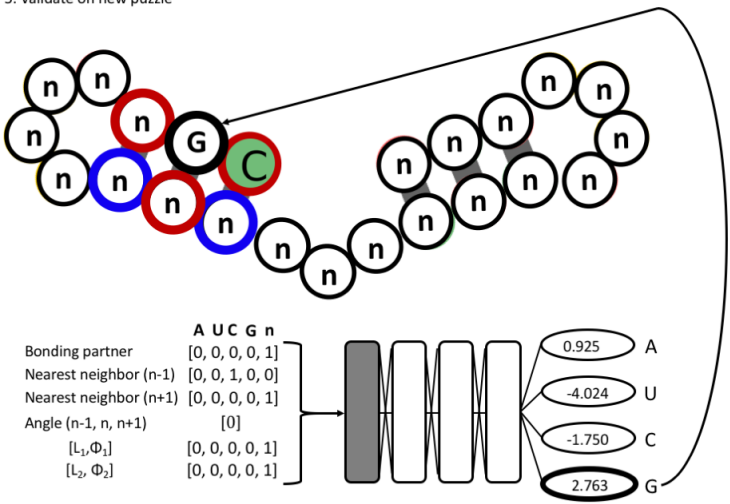

Position to be featurized
Bonding partner / nearest neighbors
Long-range features

**Figure 2:** The training and validation procedure for SentRNA consists of first selecting a puzzle, and then performing a pairwise mutual information calculation using all player solutions for that puzzle. Positions in the resulting mutual information matrix with high values are then used to define new long-range features that will be included in the model's field of vision (steps 1-3). These features are appended to the base input vector that by default has information about the bonding partner and nearest neighbors. SentRNA is then trained to reproduce a player solution at each position in the puzzle. To train the model, we use a two-part training set consisting of both the full player solution as well as a synthetic "solution trajectory" to simulate the process of solving a puzzle base-by-base starting from a blank puzzle (step 4). During validation (and subsequent testing), the model is exposed to each position in a new (initially blank) puzzle sequentially and greedily fills in bases one-by-one based on the model outputs (step 5).

Refinement algorithm:
During testing, if the initial predicted solution does not fold into the target structure, as judged by Vienna 1.8.5, we further refine this solution using an adaptive walk. We use the following refinement moves: 1) pairing two unpaired bases that should be paired in the target structure, 2) re-pairing two paired bases that should be paired, 3) unpairing two paired bases that should not be paired, and 4) G or U-U boosting,[23] two common stabilization strategies taught to beginning EteRNA players. During refinement, random trajectories of these moves are generated and applied to the initial sequence until one that folds into the target structure is reached. At any point, if an intermediate sequence is reached that folds into a structure more structurally similar to the target, which we define as the fraction of matching characters in the dot-bracket notations, the refinement trajectory is ended, and all subsequent trajectories begin from that point. Unless otherwise noted (see Results), we limited the refinement to 300 trajectories of length 30, which takes at most 90 seconds for most puzzles in the Eterna100 (~100 bases or fewer in length). By comparison, all algorithms tested in the previous benchmark by Anderson-Lee et al. were given a much longer time limit of 24 hours.[17]

      We also investigated the impact of the human prior information encoded solely in the neural network prediction itself by repeating the refinement while removing the unpairing and boosting moves (the ones that encode human strategies), allowing the refinement to only pair bases using GC, AU, or GU pairing. Interestingly, we observe that not only is proper neural network prediction critical in many cases to solving the puzzle in comparison to existing methods, some puzzles could only be solved using a combination of neural network prediction and the restricted refinement moveset consisting of only pairing moves. We found that random application of unpairing and boosting moves can in fact be detrimental in some cases by irreversibly overwriting the strategies encoded by the neural network prediction (see Results).

**Results:**
*Overall performance of SentRNA on Eterna100*
In total, SentRNA can solve 78 / 100 puzzles from the Eterna100 across all 154 models trained and tested (Table 1). Notably, of these puzzles, SentRNA can solve 47 through its initial neural network prediction alone (i.e. no refinement), already placing it ahead of or on par with 3 / 6 algorithms previously benchmarked,[17] indicating that the human-like strategies learned by

SentRNA from the *eternasolves* training set are indeed directly generalizable to more difficult puzzles and can constitute a useful prior for design.

| Puzzle | Pred | Ref | Time (s) | % Seq | Puzzle | Pred | Ref | Time (s) | % Seq |
|---|---|---|---|---|---|---|---|---|---|
| Simple Hairpin | X | X | 0.0 | 1.0 | medallion | X | X | 0.0 | 1.0 |
| Arabidopsis Thaliana 6 RNA - Difficulty Level 0 | X | X | 0.0 | 1.0 | [RNA] Repetitious Sequences 8/10 | X | X | 0.0 | 1.0 |
| Prion Pseudoknot - Difficulty Level 0 | X | X | 0.0 | 1.0 | Documenting repetitious behavior | X | X | 0.0 | 1.0 |
| Human Integrated Adenovirus - Difficulty Level 0 | | X | 0.058 | 1.0 | 7 multiloop | | X | 6.576 | 0.826 |
| The Gammaretrovirus Signal - Difficulty Level 0 | X | X | 0.0 | 1.0 | Kyurem 7 | | X | 3.363 | 0.863 |
| Saccharomyces Cerevisiae - Difficulty Level 0 | | X | 29.189 | 0.921 | JF1 | X | X | 0.0 | 1.0 |
| Fractal 2 | | X | 5.786 | 0.965 | multilooping fun | X | X | 0.0 | 1.0 |
| G-C Placement | X | X | 0.0 | 1.0 | Multiloop… | X | X | 0.0 | 1.0 |
| The Sun | | X | 13.48 | 0.915 | hard Y | | | inf | 0.0 |
| Frog Foot | X | X | 0.0 | 1.0 | Mat - Elements & Sections | | | inf | 0.0 |
| InfoRNA test 16 | X | X | 0.0 | 1.0 | Chicken feet | | X | 5.786 | 0.821 |
| Mat - Martian 2 | X | X | 0.0 | 1.0 | Bug 18 | | X | 2.76 | 0.774 |
| square | X | X | 0.0 | 1.0 | Fractal star x5 | X | X | 0.0 | 1.0 |
| Six legd turtle 2 | X | X | 0.0 | 1.0 | Crop circle 2 | | | inf | 0.0 |
| Small and Easy 6 | | X | 6.18 | 0.8 | Branching Loop | | | inf | 0.0 |
| Fractile | X | X | 0.0 | 1.0 | Bug 38 | | X | 3.7 | 0.667 |
| Six legd Turtle | X | X | 0.0 | 1.0 | Simple Single Bond | | | inf | 0.0 |
| snoRNA SNORD64 | X | X | 0.0 | 1.0 | Taraxacum officinale | X | X | 0.0 | 1.0 |
| Chalk Outline | X | X | 0.0 | 1.0 | Headless Bug on Windshield | | X | 7.696 | 0.81 |
| InfoRNA bulge test 9 | X | X | 0.0 | 1.0 | Pokeball | | X | 0.074 | 0.989 |
| Tilted Russian Cross | | X | 1.606 | 1.0 | Variation of a crop circle | | | inf | 0.0 |
| This is ACTUALLY Small And Easy 6 | X | X | 0.0 | 1.0 | Loop next to a Multiloop | | | inf | 0.0 |
| Shortie 4 | X | X | 0.0 | 1.0 | Snowflake 4 | | X | 1849.862 | 0.659 |
| Shape Test | X | X | 0.0 | 1.0 | Mat - Cuboid | | X | 0.394 | 1.0 |
| The Minitsry | X | X | 0.0 | 1.0 | Misfolded Aptamer 6 | | | inf | 0.0 |
| stickshift | X | X | 0.0 | 1.0 | Snowflake 3 | | | inf | 0.0 |
| U | X | X | 0.0 | 1.0 | Hard Y and a bit more | | | inf | 0.0 |
| Still Life (Sunflower In A Vase) | | X | 20.545 | 0.921 | Mat - Lot 2-2 B | | X | 231.217 | 0.901 |
| Quasispecies 2-2 Loop Challenge | X | X | 0.0 | 1.0 | Shapes and Energy | | | inf | 0.0 |
| Corner bulge training | X | X | 0.0 | 1.0 | Spiral of 5's | | | inf | 0.0 |
| Spiral | X | X | 0.0 | 1.0 | Campfire | | | inf | 0.0 |
| InfoRNA bulge test | X | X | 0.0 | 1.0 | Anemone | | X | 0.611 | 0.944 |
| Worm 1 | X | X | 0.0 | 1.0 | Fractal 3 | X | X | 0.0 | 1.0 |
| just down to 1 bulge | X | X | 0.0 | 1.0 | Kyurem 5 | | X | 0.852 | 0.769 |
| Iron Cross | | X | 2.06 | 0.967 | Snowflake Necklace ( or v2.0 ) | | | inf | 0.0 |
| loops and stems | X | X | 0.0 | 1.0 | Methaqualone C16H14N2O Structural Representation | | | inf | 0.0 |
| Water Strider | | X | 4.863 | 0.942 | Cat's Toy 2 | | | inf | 0.0 |
| The Turtle(s) Move(s) | X | X | 0.0 | 1.0 | Zigzag Semicircle | | | inf | 0.0 |
| Adenine | X | X | 0.0 | 1.0 | Short String 4 | | | inf | 0.0 |
| Tripod5 | | X | 0.331 | 1.0 | Gladius | | | inf | 0.0 |
| Shortie 6 | X | X | 0.0 | 1.0 | Thunderbolt | | | inf | 0.0 |
| Runner | | X | 7.089 | 0.931 | Mutated chicken feet | | X | 415.076 | 0.7 |
| Recoil | X | X | 0.0 | 1.0 | Chicken Tracks | | X | 0.175 | 0.879 |
| [CloudBeta] An Arm and a Leg 1.0 | X | X | 0.0 | 1.0 | Looking Back Again | | | inf | 0.0 |
| [CloudBeta] 5 Adjacent Stack Multi-Branch Loop | | X | 1.034 | 0.968 | Multilooping 6 | | | inf | 0.0 |
| Triple Y | | X | 8.891 | 0.794 | Cesspool | | | inf | 0.0 |
| Misfolded Aptamer | | X | 6.659 | 0.8 | Hoglafractal | | | inf | 0.0 |
| Flower power | | X | 6.957 | 0.843 | Bullseye | | X | 207.149 | 0.673 |
| Kudzu | | X | 5.246 | 0.861 | Shooting Star | | | inf | 0.0 |
| "1,2,3and4bulges" | X | X | 0.0 | 1.0 | Teslagon | | | inf | 0.0 |

**Table 1**: Performance of the SentRNA ensemble of 154 trained models on the full set of Eterna100 puzzles. An "X" in the "Pred" column means that a model could predict a correct solution to the puzzle without refinement, and an "X" in the "Ref" column means the model was able to solve the puzzle after refinement. The number in the "Times" column represents the minimum refinement time across all models that was necessary to solve the puzzle, in seconds, and "% Seq" corresponds to the fraction sequence identity between the neural network prediction and the refined solution for that model. These results were generated using default training and refinement parameters: 50 randomly selected player solutions for training, and 300 adaptive walk trajectories of length 30 using the full moveset of pairing, unpairing, and boosting moves.

We also more explicitly tested the contribution of the prior information encoded in the neural network prediction by repeating the refinement process while allowing SentRNA to only perform pairing moves (GC, AU, GU), making it equivalent to the original adaptive walk procedure used in RNAInverse.[6] By restricting the moveset in this way, we ensure that any potential advantages contributed by human prior information can only be contributed by the neural network prediction. Put another way, we place SentRNA's refinement algorithm on par with the stochastic search algorithms employed by methods such as NUPACK and MODENA in terms of contributable information content, i.e. there is no change to the sequence that SentRNA's restricted moveset can make that cannot be made by these other search procedures. This approach allows us to more directly measure the potential advantages of SentRNA's neural network initialization over the initialization methods of other algorithms. We performed two rounds of refinement for all the models: 1) using a moveset consisting of all pairing moves (GC, AU, and GU) (Table 2) and 2) using a more aggressive, purely GC mutation moveset (Table 3).

Remarkably, we find that SentRNA's performance is comparable to when using the full moveset. SentRNA can solve 74 / 100 puzzles using neural network prediction + GC pairing and 72 / 100 using neural network prediction + GC/AU/GU pairing, indicating that the neural network initialization by itself contributes significantly to SentRNA's performance. We summarize the overall performance of the different refinement movesets in Table 4.

| Puzzle | Pred | Ref | Time (s) | % Seq | Puzzle | Pred | Ref | Time (s) | % Seq |
|---|---|---|---|---|---|---|---|---|---|
| Simple Hairpin | X | X | 0.0 | 1.0 | medallion | X | X | 0.0 | 1.0 |
| Arabidopsis Thaliana 6 RNA - Difficulty Level 0 | X | X | 0.0 | 1.0 | [RNA] Repetitious Sequences 8/10 | X | X | 0.0 | 1.0 |
| Prion Pseudoknot - Difficulty Level 0 | X | X | 0.0 | 1.0 | Documenting repetitious behavior | X | X | 0.0 | 1.0 |
| Human Integrated Adenovirus - Difficulty Level 0 | | X | 0.065 | 1.0 | 7 multiloop | | X | 9.63 | 0.826 |
| The Gammaretrovirus Signal - Difficulty Level 0 | X | X | 0.0 | 1.0 | Kyurem 7 | | X | 39.014 | 0.706 |
| Saccharomyces Cerevisiae - Difficulty Level 0 | | X | 362.555 | 0.712 | JF1 | X | X | 0.0 | 1.0 |
| Fractal 2 | | X | 4.093 | 0.958 | multilooping fun | | X | 0.107 | 1.0 |
| G-C Placement | X | X | 0.0 | 1.0 | Multiloop… | X | X | 0.0 | 1.0 |
| The Sun | | X | 180.932 | 0.845 | hard Y | | | inf | 0.0 |
| Frog Foot | X | X | 0.0 | 1.0 | Mat - Elements & Sections | | | inf | 0.0 |
| InfoRNA test 16 | X | X | 0.0 | 1.0 | Chicken feet | | X | 627.204 | 0.821 |
| Mat - Martian 2 | X | X | 0.0 | 1.0 | Bug 18 | | X | 5.638 | 0.811 |
| square | X | X | 0.0 | 1.0 | Fractal star x5 | X | X | 0.0 | 1.0 |
| Six legd turtle 2 | X | X | 0.0 | 1.0 | Crop circle 2 | | | inf | 0.0 |
| Small and Easy 6 | | X | 13.288 | 0.733 | Branching Loop | | | inf | 0.0 |
| Fractile | X | X | 0.0 | 1.0 | Bug 38 | | X | 18.495 | 0.778 |
| Six legd Turtle | X | X | 0.0 | 1.0 | Simple Single Bond | | | inf | 0.0 |
| snoRNA SNORD64 | X | X | 0.0 | 1.0 | Taraxacum officinale | X | X | 0.0 | 1.0 |
| Chalk Outline | X | X | 0.0 | 1.0 | Headless Bug on Windshield | | X | 413.93 | 0.93 |
| InfoRNA bulge test 9 | X | X | 0.0 | 1.0 | Pokeball | | X | 1.832 | 0.973 |
| Tilted Russian Cross | | X | 0.391 | 1.0 | Variation of a crop circle | | | inf | 0.0 |
| This is ACTUALLY Small And Easy 6 | X | X | 0.0 | 1.0 | Loop next to a Multiloop | | | inf | 0.0 |
| Shortie 4 | X | X | 0.0 | 1.0 | Snowflake 4 | | | inf | 0.0 |
| Shape Test | X | X | 0.0 | 1.0 | Mat - Cuboid | X | X | 0.0 | 1.0 |
| The Minitsry | | X | 0.193 | 1.0 | Misfolded Aptamer 6 | | | inf | 0.0 |
| stickshift | X | X | 0.0 | 1.0 | Snowflake 3 | | | inf | 0.0 |
| U | X | X | 0.0 | 1.0 | Hard Y and a bit more | | | inf | 0.0 |
| Still Life (Sunflower In A Vase) | | X | 27.408 | 0.854 | Mat - Lot 2-2 B | | X | 8276.732 | 0.704 |
| Quasispecies 2-2 Loop Challenge | X | X | 0.0 | 1.0 | Shapes and Energy | | X | 255.671 | 0.743 |
| Corner bulge training | X | X | 0.0 | 1.0 | Spiral of 5's | | | inf | 0.0 |
| Spiral | X | X | 0.0 | 1.0 | Campfire | | | inf | 0.0 |
| InfoRNA bulge test | X | X | 0.0 | 1.0 | Anemone | | X | 5.253 | 0.921 |
| Worm 1 | X | X | 0.0 | 1.0 | Fractal 3 | X | X | 0.0 | 1.0 |
| just down to 1 bulge | X | X | 0.0 | 1.0 | Kyurem 5 | | X | 1497.968 | 0.788 |
| Iron Cross | | X | 5.591 | 0.87 | Snowflake Necklace ( or v2.0 ) | | | inf | 0.0 |
| loops and stems | | X | 0.09 | 1.0 | Methaqualone C16H14N2O Structural Representation | | | inf | 0.0 |
| Water Strider | | X | 295.483 | 0.796 | Cat's Toy 2 | | | inf | 0.0 |
| The Turtle(s) Move(s) | X | X | 0.0 | 1.0 | Zigzag Semicircle | | | inf | 0.0 |
| Adenine | X | X | 0.0 | 1.0 | Short String 4 | | | inf | 0.0 |
| Tripod5 | | X | 0.178 | 1.0 | Gladius | | | inf | 0.0 |
| Shortie 6 | X | X | 0.0 | 1.0 | Thunderbolt | | | inf | 0.0 |
| Runner | | X | 7.097 | 0.948 | Mutated chicken feet | | | inf | 0.0 |
| Recoil | X | X | 0.0 | 1.0 | Chicken Tracks | | X | 66.241 | 0.853 |
| [CloudBeta] An Arm and a Leg 1.0 | X | X | 0.0 | 1.0 | Looking Back Again | | | inf | 0.0 |
| [CloudBeta] 5 Adjacent Stack Multi-Branch Loop | | X | 0.859 | 0.937 | Multilooping 6 | | | inf | 0.0 |
| Triple Y | | X | 40.128 | 0.845 | Cesspool | | | inf | 0.0 |
| Misfolded Aptamer | | X | 299.134 | 0.96 | Hoglafractal | | | inf | 0.0 |
| Flower power | | X | 33.138 | 0.765 | Bullseye | | | inf | 0.0 |
| Kudzu | | X | 26.276 | 0.941 | Shooting Star | | | inf | 0.0 |
| "1,2,3and4bulges" | X | X | 0.0 | 1.0 | Teslagon | | | inf | 0.0 |

**Table 2:** Performance of the SentRNA ensemble of 154 trained models on the Eterna100 when restricting the refinement moveset to only GC, AU, and GU pairing moves. An "X" in the "Pred" column means that a model could predict a correct solution to the puzzle without refinement, and an "X" in the "Ref" column means the model was able to solve the puzzle after refinement. The number in the "Times" column represents the minimum refinement time across all models that was necessary to solve the puzzle, in seconds, and "% Seq" corresponds to the fraction sequence identity between the neural network prediction and the refined solution for that model. These results were generated using default training and refinement parameters: 50 randomly selected player solutions for training, and 300 adaptive walk trajectories of length 30 using the restricted moveset.

| Puzzle | Pred | Ref | Time (s) | % Seq | Puzzle | Pred | Ref | Time (s) | % Seq |
|---|---|---|---|---|---|---|---|---|---|
| Simple Hairpin | X | X | 0.0 | 1.0 | medallion | X | X | 0.0 | 1.0 |
| Arabidopsis Thaliana 6 RNA - Difficulty Level 0 | X | X | 0.0 | 1.0 | [RNA] Repetitious Sequences 8/10 | X | X | 0.0 | 1.0 |
| Prion Pseudoknot - Difficulty Level 0 | X | X | 0.0 | 1.0 | Documenting repetitious behavior | X | X | 0.0 | 1.0 |
| Human Integrated Adenovirus - Difficulty Level 0 | | X | 0.058 | 1.0 | 7 multiloop | | X | 6.576 | 0.826 |
| The Gammaretrovirus Signal - Difficulty Level 0 | X | X | 0.0 | 1.0 | Kyurem 7 | | X | 3.363 | 0.863 |
| Saccharomyces Cerevisiae - Difficulty Level 0 | | X | 29.189 | 0.921 | JF1 | X | X | 0.0 | 1.0 |
| Fractal 2 | | X | 5.786 | 0.965 | multilooping fun | X | X | 0.0 | 1.0 |
| G-C Placement | X | X | 0.0 | 1.0 | Multiloop… | X | X | 0.0 | 1.0 |
| The Sun | | X | 13.48 | 0.915 | hard Y | | | inf | 0.0 |
| Frog Foot | X | X | 0.0 | 1.0 | Mat - Elements & Sections | | | inf | 0.0 |
| InfoRNA test 16 | X | X | 0.0 | 1.0 | Chicken feet | | X | 5.786 | 0.821 |
| Mat - Martian 2 | X | X | 0.0 | 1.0 | Bug 18 | | X | 2.76 | 0.774 |
| square | X | X | 0.0 | 1.0 | Fractal star x5 | X | X | 0.0 | 1.0 |
| Six legd turtle 2 | X | X | 0.0 | 1.0 | Crop circle 2 | | | inf | 0.0 |
| Small and Easy 6 | | X | 6.18 | 0.8 | Branching Loop | | | inf | 0.0 |
| Fractile | X | X | 0.0 | 1.0 | Bug 38 | | X | 3.7 | 0.667 |
| Six legd Turtle | X | X | 0.0 | 1.0 | Simple Single Bond | | | inf | 0.0 |
| snoRNA SNORD64 | X | X | 0.0 | 1.0 | Taraxacum officinale | X | X | 0.0 | 1.0 |
| Chalk Outline | X | X | 0.0 | 1.0 | Headless Bug on Windshield | | X | 7.696 | 0.81 |
| InfoRNA bulge test 9 | X | X | 0.0 | 1.0 | Pokeball | | X | 0.074 | 0.989 |
| Tilted Russian Cross | | X | 1.606 | 1.0 | Variation of a crop circle | | | inf | 0.0 |
| This is ACTUALLY Small And Easy 6 | X | X | 0.0 | 1.0 | Loop next to a Multiloop | | | inf | 0.0 |
| Shortie 4 | X | X | 0.0 | 1.0 | Snowflake 4 | | X | 1849.862 | 0.659 |
| Shape Test | X | X | 0.0 | 1.0 | Mat - Cuboid | | X | 0.394 | 1.0 |
| The Minitsry | | X | 0.0 | 1.0 | Misfolded Aptamer 6 | | | inf | 0.0 |
| stickshift | X | X | 0.0 | 1.0 | Snowflake 3 | | | inf | 0.0 |
| U | X | | 0.0 | 1.0 | Hard Y and a bit more | | | inf | 0.0 |
| Still Life (Sunflower In A Vase) | | X | 20.545 | 0.921 | Mat - Lot 2-2 B | | X | 231.217 | 0.901 |
| Quasispecies 2-2 Loop Challenge | X | X | 0.0 | 1.0 | Shapes and Energy | | | inf | 0.0 |
| Corner bulge training | X | X | 0.0 | 1.0 | Spiral of 5's | | | inf | 0.0 |
| Spiral | X | X | 0.0 | 1.0 | Campfire | | | inf | 0.0 |
| InfoRNA bulge test | X | X | 0.0 | 1.0 | Anemone | | X | 0.611 | 0.944 |
| Worm 1 | X | X | 0.0 | 1.0 | Fractal 3 | X | X | 0.0 | 1.0 |
| just down to 1 bulge | X | X | 0.0 | 1.0 | Kyurem 5 | | X | 0.852 | 0.769 |
| Iron Cross | | X | 2.06 | 0.967 | Snowflake Necklace ( or v2.0 ) | | | inf | 0.0 |
| loops and stems | X | X | 0.0 | 1.0 | Methaqualone C16H14N2O Structural Representation | | | inf | 0.0 |
| Water Strider | | X | 4.863 | 0.942 | Cat's Toy 2 | | | inf | 0.0 |
| The Turtle(s) Move(s) | X | X | 0.0 | 1.0 | Zigzag Semicircle | | | inf | 0.0 |
| Adenine | X | X | 0.0 | 1.0 | Short String 4 | | | inf | 0.0 |
| Tripod5 | | X | 0.331 | 1.0 | Gladius | | | inf | 0.0 |
| Shortie 6 | X | X | 0.0 | 1.0 | Thunderbolt | | | inf | 0.0 |
| Runner | | X | 7.089 | 0.931 | Mutated chicken feet | | X | 415.076 | 0.7 |
| Recoil | X | X | 0.0 | 1.0 | Chicken Tracks | | X | 0.175 | 0.879 |
| [CloudBeta] An Arm and a Leg 1.0 | X | X | 0.0 | 1.0 | Looking Back Again | | | inf | 0.0 |
| [CloudBeta] 5 Adjacent Stack Multi-Branch Loop | | X | 1.034 | 0.968 | Multilooping 6 | | | inf | 0.0 |
| Triple Y | | X | 8.891 | 0.794 | Cesspool | | | inf | 0.0 |
| Misfolded Aptamer | | X | 6.659 | 0.8 | Hoglafractal | | | inf | 0.0 |
| Flower power | | X | 6.957 | 0.843 | Bullseye | | X | 207.149 | 0.673 |
| Kudzu | | X | 5.246 | 0.861 | Shooting Star | | | inf | 0.0 |
| "1,2,3and4bulges" | X | X | 0.0 | 1.0 | Teslagon | | | inf | 0.0 |

**Table 3:** Performance of the SentRNA ensemble of 154 trained models on the Eterna100 when restricting the refinement moveset to only GC pairing moves. An "X" in the "Pred" column means that a model could predict a correct solution to the puzzle without refinement, and an "X" in the "Ref" column means the model was able to solve the puzzle after refinement. The number in the "Times" column represents the minimum refinement time across all models that was necessary to solve the puzzle, in seconds, and "% Seq" corresponds to the fraction sequence identity between the neural network prediction and the refined solution for that model. These results were generated using default training and refinement parameters: 50 randomly selected player solutions for training, and 300 adaptive walk trajectories of length 30 using the restricted moveset.

| Puzzle | Pred | Full | All pair | GC pair | Puzzle | Pred | Full | All pair | GC pair |
|---|---|---|---|---|---|---|---|---|---|
| Simple Hairpin | X | X | X | X | medallion | X | X | X | X |
| Arabidopsis Thaliana 6 RNA - Difficulty Level 0 | X | X | X | X | [RNA] Repetitious Sequences 8/10 | X | X | X | X |
| Prion Pseudoknot - Difficulty Level 0 | X | X | X | X | Documenting repetitious behavior | X | X | X | X |
| Human Integrated Adenovirus - Difficulty Level 0 | X | X | X | X | 7 multiloop | | X | X | X |
| The Gammaretrovirus Signal - Difficulty Level 0 | X | X | X | X | Kyurem 7 | | X | X | X |
| Saccharomyces Cerevisiae - Difficulty Level 0 | | X | X | X | JF1 | X | X | X | X |
| Fractal 2 | | X | X | X | multilooping fun | X | X | X | X |
| G-C Placement | X | X | X | X | Multiloop... | X | X | X | X |
| The Sun | | X | X | X | hard Y | | X | | |
| Frog Foot | X | X | X | X | Mat - Elements & Sections | | | | |
| InfoRNA test 16 | X | X | X | X | Chicken feet | | X | X | X |
| Mat - Martian 2 | X | X | X | X | Bug 18 | | X | X | X |
| square | X | X | X | X | Fractal star x5 | X | X | X | X |
| Six legd turtle 2 | X | X | X | X | Crop circle 2 | | | | |
| Small and Easy 6 | | X | X | X | Branching Loop | | X | | |
| Fractile | X | X | X | X | Bug 38 | | X | X | X |
| Six legd Turtle | X | X | X | X | Simple Single Bond | | X | | |
| snoRNA SNORD64 | X | X | X | X | Taraxacum officinale | X | X | X | X |
| Chalk Outline | X | X | X | X | Headless Bug on Windshield | | X | X | X |
| InfoRNA bulge test 9 | X | X | X | X | Pokeball | | X | X | X |
| Tilted Russian Cross | X | X | X | X | Variation of a crop circle | | X | | |
| This is ACTUALLY Small And Easy 6 | X | X | X | X | Loop next to a Multiloop | | X | | |
| Shortie 4 | X | X | X | X | Snowflake 4 | | | | X |
| Shape Test | X | X | X | X | Mat - Cuboid | X | X | X | X |
| The Minitsry | X | X | X | X | Misfolded Aptamer 6 | | X | | |
| stickshift | X | X | X | X | Snowflake 3 | | | | |
| U | X | X | X | X | Hard Y and a bit more | | | | |
| Still Life (Sunflower In A Vase) | | X | X | X | Mat - Lot 2-2 B | | | X | X |
| Quasispecies 2-2 Loop Challenge | X | X | X | X | Shapes and Energy | | X | X | |
| Corner bulge training | X | X | X | X | Spiral of 5's | | X | | |
| Spiral | X | X | X | X | Campfire | | | | |
| InfoRNA bulge test | X | X | X | X | Anemone | | X | X | X |
| Worm 1 | X | X | X | X | Fractal 3 | X | X | X | X |
| just down to 1 bulge | X | X | X | X | Kyurem 5 | | | X | X |
| Iron Cross | | X | X | X | Snowflake Necklace ( or v2.0 ) | | | | |
| loops and stems | X | X | X | X | Methaqualone C16H14N2O Structural Representation | | | | |
| Water Strider | | X | X | X | Cat's Toy 2 | | | | |
| The Turtle(s) Move(s) | X | X | X | X | Zigzag Semicircle | | | | |
| Adenine | X | X | X | X | Short String 4 | | | | |
| Tripod5 | X | X | X | X | Gladius | | | | |
| Shortie 6 | X | X | X | X | Thunderbolt | | X | | |
| Runner | | X | X | X | Mutated chicken feet | | | | X |
| Recoil | X | X | X | X | Chicken Tracks | X | X | X | X |
| [CloudBeta] An Arm and a Leg 1.0 | X | X | X | X | Looking Back Again | | | | |
| [CloudBeta] 5 Adjacent Stack Multi-Branch Loop | | X | X | X | Multilooping 6 | | | | |
| Triple Y | | X | X | X | Cesspool | | | | |
| Misfolded Aptamer | | X | X | X | Hoglafractal | | | | |
| Flower power | | X | X | X | Bullseye | | | | X |
| Kudzu | | X | X | X | Shooting Star | | | | |
| "1,2,3and4bulges" | X | X | X | X | Teslagon | | | | |

**Table 4:** Overall comparison of SentRNA's performance given different refinement movesets. Solved puzzles are denoted with an "X". The "Pred" column refers to only neural network prediction. "Full" refers to puzzles solved using the full refinement moveset of pairing, unpairing, and boosting moves. "All pair" refers to neural network prediction + all possible pairing moves, GC/AU/GU, and "GC pair" refers to neural network prediction + GC pairing moves only.

*SentRNA can learn critical solution strategies and apply them through its neural network prediction*

SentRNA can solve 47 / 100 puzzles using neural network refinement alone, suggesting that the human strategies learned during training are indeed generalizable to more difficult targets. Specifically, we find that for many of the puzzles solvable purely through neural network prediction, SentRNA shows clear human-like signatures in its predicted sequences. For example, for the puzzle *"1,2,3and4bulges"* (Figure 4, left), the most difficult structural component to stabilize is the unstable length-1 stem attached to a 4-loop. It is necessary to boost this 4-loop with a G at the N-terminal end of the loop to solve the puzzle. We observe that SentRNA can apply this strategy directly through its neural network prediction, allowing it to solve the puzzle without any needed refinement.

Similarly, for the puzzle *Mat – Cuboid*, a key stabilization strategy for this puzzle is a specific boosting strategy for the 2-2 loops invented by the EteRNA community called a "UGUG superboost" and involves mutating all four bases of each 2-2 loop to UGUG (Figure 4, right). We see that SentRNA can learn this strategy through the training data and generalize it to a much more difficult puzzle in the Eterna100.

Through these two examples, we see that by if SentRNA is provided the appropriate training data, it can learn human-like strategies and apply them to help solve new, more difficult targets. This potentially gives SentRNA unique advantages over other available inverse design algorithms that do not have knowledge of these strategies when solving puzzles rich in 4-loops or 2-2 loops.

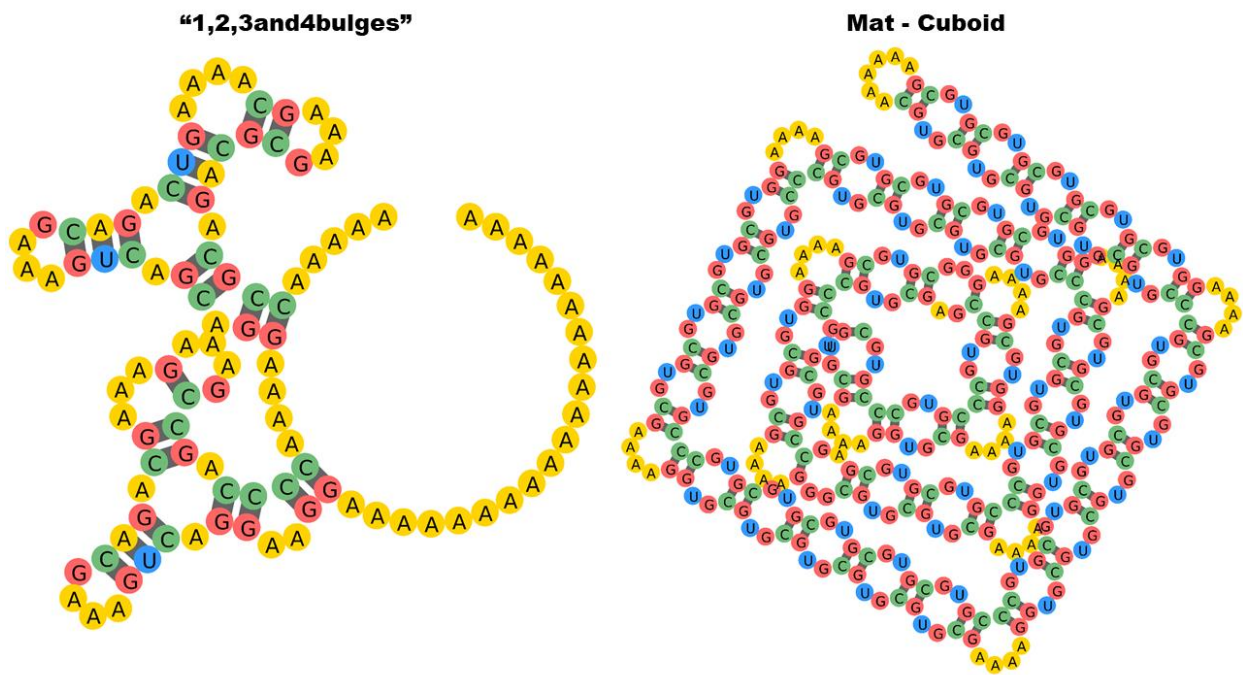

**Figure 4:** SentRNA can learn human-like strategies such as 4-loop boosting and the UGUG superboost for 2-2 loops that enables it to solve puzzles such as *"1,2,3and4bulges"* and *Mat – Cuboid*.

*Prior information included in the refinement moveset compensates for deficiencies in neural network training*

We found that some puzzles in the Eterna100 necessitated the use of the full refinement moveset, which includes pairing, unpairing, and boosting moves. If SentRNA is unable to learn the necessary stabilization strategies for a specific puzzle, or if the strategies learned from training are not perfectly generalizable to a new puzzle, the human prior information encoded in the unpairing and boosting moves can be critical to reaching a valid solution. For example, the puzzle *hard Y* (Figure 5, left) contains an unusual structural element consisting of two adjacent length-1 bulges, named the "zigzag" by the EteRNA community. A key step in stabilizing this structure is mutating one of the unpaired bulge bases to C. However, because the zigzag is a rarely seen structural element, SentRNA was unable to learn the proper stabilization strategy through the training data. However, the presence of the unpairing move allows SentRNA to sample the necessary stabilization move during its refinement trajectory and solve the puzzle (Figure 5, right).

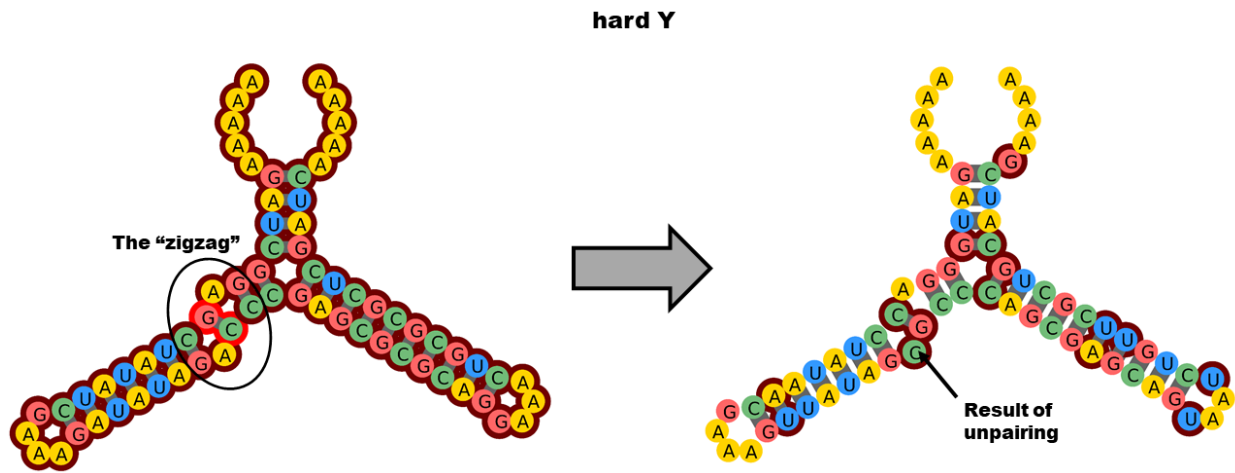

**Figure 5**: The puzzle *hard Y* contains a rarely seen structural element of two adjacent length-1 bulges called a zigzag. We observed it was necessary to mutate one of these bulge bases to a C to stabilize the zigzag and solve the puzzle. SentRNA was unable to learn this strategy during training, necessitating the use of an unpairing move during refinement to perform the C-mutation.

As another example, although SentRNA can often apply productive boosting moves with its neural network prediction, such as for *"1,2,3and4bulges"*, we also observed that SentRNA's knowledge of boosting is incomplete. SentRNA can readily learn how to boost using G-mutations due to the ubiquitous use of this strategy in the EteRNA community. However, U-U boosting, a more advanced strategy, is less well-represented in the training set, and we observed that SentRNA had difficulty learning it via its neural network. For example, in the puzzle *Misfolded Aptamer 6*, SentRNA attempts to boost the internal loops using G-mutations only. However, as the refinement reveals, a U-U boost to the 2-5 internal loop appears necessary to solve the puzzle. Thus, in this situation, refinement is necessary to compensate for the lack of prior knowledge in SentRNA's neural network (Figure 6).

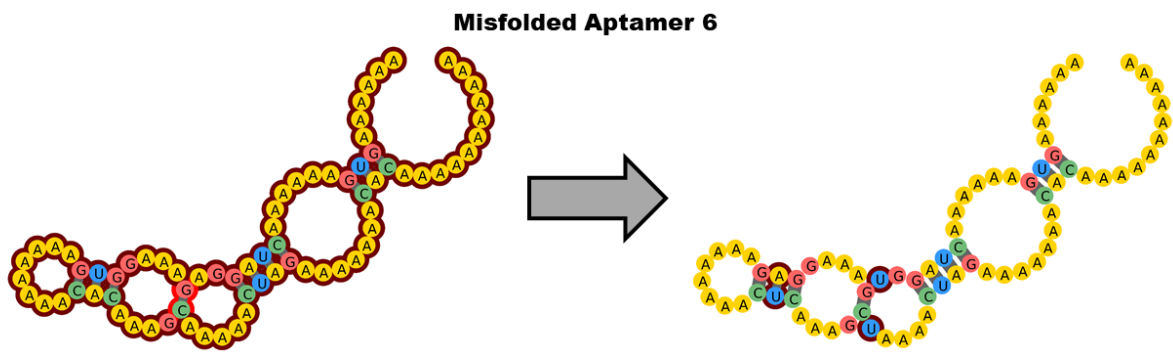

**Figure 6:** SentRNA applies boosts to the 4-4 and 2-5 internal loops of *Misfolded Aptamer 6* by mutating three separate bases on these loops to G, but we observed that a critical stabilization move that SentRNA did not learn was an additional U-U boost at the 2-5 internal loop. This move needed to be sampled during refinement to solve the puzzle.

*Prior information included in the refinement moveset can potentially harm SentRNA's performance*

Previously, we showed that the human prior information encoded in the unpairing and boosting moves was necessary to solve some puzzles by making up for deficiencies in SentRNA's neural network training. Interestingly, however, we observe that the opposite can also be true, and that some puzzles are only solvable using neural network prediction and a restricted moveset of only pairing moves. We found that the prior information encoded in the refinement moveset, when applied in a random, undirected manner, can sometimes cause irreversible damage to an intelligently initialized sequence and make it impossible to solve certain puzzles. For example, the puzzle *Mat – Lot 2-2 B* represents a more difficult version of *Mat – Cuboid*, in which the UGUG superboost for the 2-2 loop is a necessary part of the solution strategy. Including boosting and unpairing moves into the refinement moveset can irreversibly damage a sequence that is initialized by the neural network and contains this strategy, making it impossible to solve the puzzle. We observed that the only way *Mat – Lot* 2-2 B could be solved is through a combination of neural network refinement and a restricted moveset of only pairing moves, as this fully preserves the UGUG superboost (Figure 7).

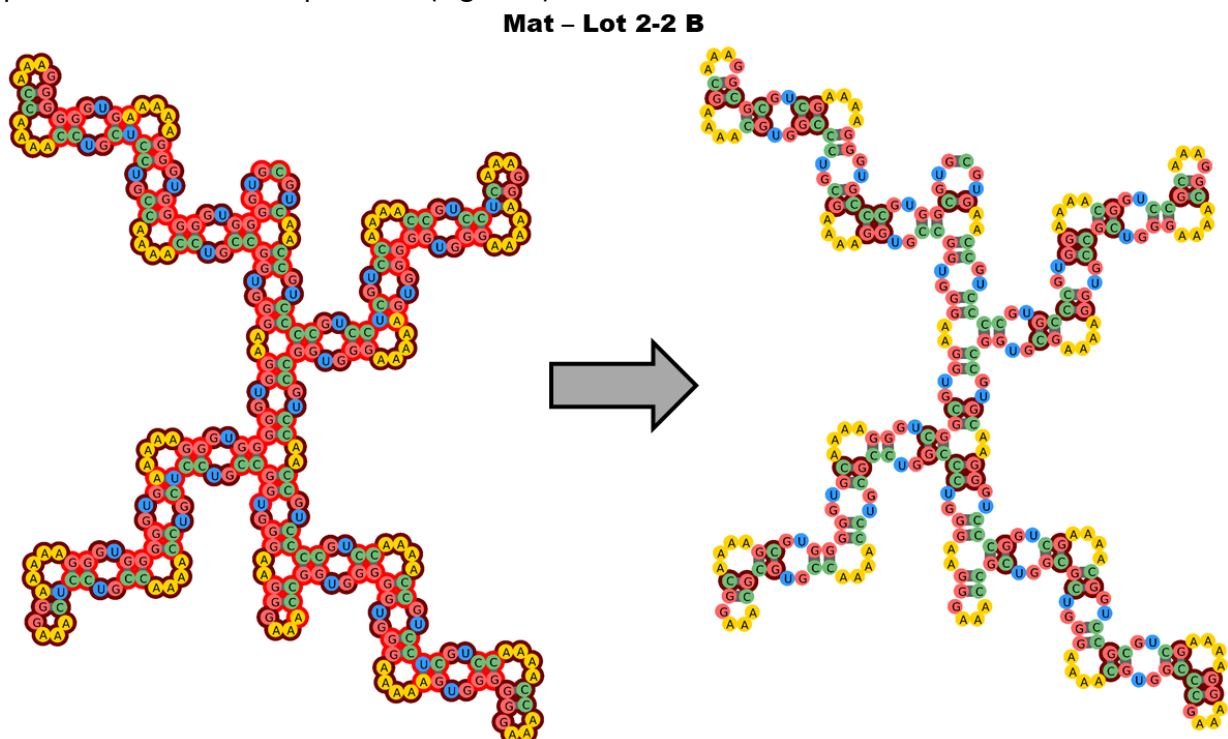

**Figure 7:** SentRNA could only solve *Mat – Lot 2-2 B* using a combination of neural network prediction to initialize the sequence with the UGUG superboost strategy, coupled with refinement using only pair moves. Allowing for unpairing and boosting moves can easily and irreversibly contaminate the UGUG superboosts and prevent SentRNA from solving the puzzle.

Another way in which unpairing and boosting moves can harm SentRNA's performance is by needlessly increasing the refinement search space such that productive moves are sampled less often. For puzzles that require only pairing moves to solve, including unpairing and boosting moves in the refinement moveset can significantly slow the refinement process, to the point in which SentRNA cannot sample a valid solution before the refinement timeout. For example, the puzzle *Bullseye* only requires GC pair mutations to solve beginning from the neural network prediction, and in fact was only solved using a combination of neural network prediction + GC pairing moves (Figure 8). The likely reason for this is due to its large size, including unnecessary moves into the moveset (AU/GU pairing, unpairing, and boosting) will substantially increase the required time to sample these productive GC mutations, and SentRNA as a result cannot solve the puzzle with anything more extensive than a GC pairing moveset.

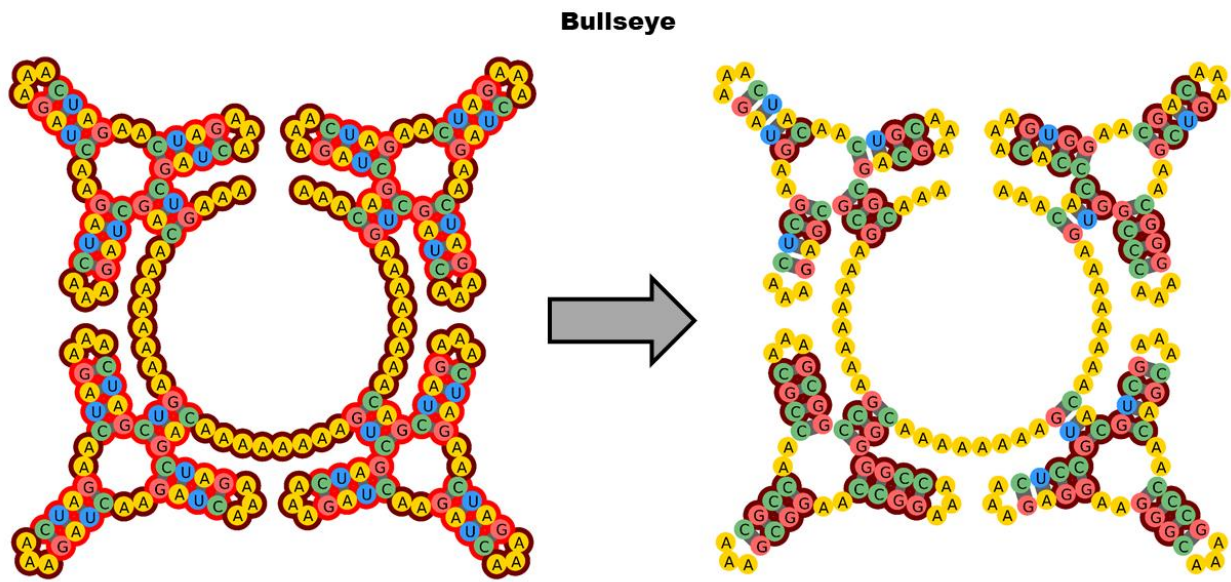

**Figure 8:** SentRNA can solve *Bullseye* using only a combination of neural network prediction and GC pairing moves during refinement. Incorporation of other move types substantially increases the refinement search space and makes it more difficult to sample the productive GC mutation moves, leading to a prohibitively long refinement time.

*Summary of SentRNA's performance on the Eterna100 versus other methods*

We summarize the performance of SentRNA on the Eterna100 vs. the 6 other methods previously benchmarked by Anderson-Lee et al in Table 5.[17] We observe that even with a restricted refinement moveset of only pairing moves, which places SentRNA's refinement algorithm on par with the stochastic search algorithms of other methods in terms of information content, SentRNA still significantly outperforms the top-performing methods MODENA, INFO-RNA, and NUPACK. These results indicate that the human prior information encoded in SentRNA's neural network prediction results in a more useful initialization compared to these other methods for a larger number of puzzles. We see this explicitly for puzzles such as *Mat – Lot 2-2 B*, where the human-developed UGUG superboost strategy is necessary for solving the puzzle, and hence this puzzle is solvable by only SentRNA and none of the other algorithms.

| Method | Number of puzzles solved |
|---|---|
| SentRNA, NN only | 47 / 100 |
| SentRNA, NN + full moveset | **78 / 100** |
| SentRNA, NN + GC pairing | 74 / 100 |
| SentRNA, NN + all pairing | 72 / 100 |
| RNAinverse | 28 / 100 |
| RNA-SSD | 27 / 100 |
| DSS-Opt | 47 / 100 |
| NUPACK | 48 / 100 |
| INFO-RNA | 50 / 100 |
| MODENA | 54 / 100 |

**Table 5:** Summary of performance of SentRNA vs. the 6 other algorithms benchmarked on the Eterna100 by Anderson-Lee et al. We see that even with a restricted moveset of only pairing moves, SentRNA still significantly outperforms other methods, indicating that the neural network prediction is an overall more useful initialization scheme than those of other algorithms for the objective of solving the Eterna100.

*SentRNA achieves state-of-the-art results on an independent, non-EteRNA test set*
Although SentRNA's state-of-the-art performance on the Eterna100 is promising, there remains the possibility that because we are both training and testing on EteRNA puzzles, SentRNA is simply overfitting to EteRNA, and may not be useful for designing sequences for non-EteRNA targets. To address this possibility, we further tested our ensemble of 154 trained models on a set of 63 non-EteRNA, experimentally synthesized targets that Garcia-Martin et al. recently used to benchmark a set of 10 inverse folding algorithms.[19] To our knowledge, this is the most recent and comprehensive benchmark of current state-of-the-art methods.

Remarkably, we find that the prior information learned from training on EteRNA puzzles is indeed generalizable to these non-EteRNA structures. We find that SentRNA can solve 46 / 63 targets using neural network prediction alone and 57 / 63 using neural network in combination with the full refinement moveset. As such, the neural network prediction on its own is sufficient to surpass 8 / 10 methods previously benchmarked, competitive with RNA-SSD (47 / 63) and only clearly worse than ERD (54 / 63). If we allow for a 10-minute refinement using the full refinement moveset (the time limit allowed by Garcia-Martin et al. for the other methods), SentRNA surpasses all previous methods (Table 6).

However, we also observed that using the restricted movset of either GC pairing moves or GC/AU/GU pairing moves allows SentRNA to solve only 53 / 63 targets, slightly worse than ERD. Therefore, it appears that while the contribution of the neural network initialization is sufficient to allow SentRNA to remain competitive with the current state-of-the-art for this test set, it does not grant a decisive advantage. In fact, ERD overall can solve one more target than SentRNA, suggesting that ERD's initialization using natural sequences from a known, biological database may be more useful than SentRNA's initialization that is learned from EteRNA puzzles for this test set. Thus, unlike the Eterna100, it appears that the additional prior information encoded in the unpairing and boosting moves is necessary in this case for SentRNA to surpass the previous state-of-the-art.

We also compared the GC content of solutions generated by SentRNA versus the other methods, averaged over all models (Table 6). We observe that the solutions generated by the NN, NN + all pairing moves, and NN + the full refinement moveset have a mean GC content of 56-58%, computed over all models and targets, while NN + GC pairing is slightly higher at 61%. This level of GC content is similar to many of the other methods (RNAfbinv, Frnakenstein, RNAInverse, NUPACK, MODENA, RNAiFold 2.0, and ERD), which on average generate sequences with a roughly 50% GC content. However, if we consider results from individual models separately, we see that depending on the specific model tested, the variation in GC content can vary dramatically given the same target (Table 7). For instance, for the target RF00008.11 from the Rfam databse, SentRNA generated solutions using its NN which range from 20% to 100% GC content, which is a consequence of the fact that models trained using different training sets can employ significantly different solution strategies. These results suggest that training an ensemble of models and then using different subsets of these models for testing can provide the user a degree of control over the amount of GC content present in a predicted solution.

| Method | Number of puzzles solved | GC content (%) |
| --- | --- | --- |
| SentRNA, NN only | 46 / 63 | 58 ± 22 |
| **SentRNA, NN + full moveset** | **57 / 63** | 57 ± 19 |
| SentRNA, NN + all pairing | 53 / 63 | 56 ± 19 |
| SentRNA, NN + GC pairing | 53 / 63 | 61 ± 21 |
| RNAfbinv | 0 / 63 | 51 |
| IncaRNAtion | 6 / 63 | 71 |
| Frnakenstein | 19 / 63 | 49 |
| RNAInverse | 20 / 63 | 49 |
| NUPACK | 29 / 63 | 57 |
| MODENA | 32 / 63 | 50 |
| RNAiFold 2.0 | 41 / 63 | 57 |
| INFO-RNA | 45 / 63 | 72 |
| RNA-SSD | 47 / 63 | 36 |
| ERD | 54 / 63 | 55 |

**Table 6:** Comparison of SentRNA vs. 10 other inverse folding algorithms on the 63-structure test set of Garcia-Martin et al.[19] SentRNA's neural network alone can solve 46 / 63 structures, placing it ahead of 8 / 10 algorithms. With neural network prediction + refinement using the full moveset, and limiting the refinement to 10 minutes per target, SentRNA achieves state-of-the-art performance and solves 57 / 63 targets, outperforming all other algorithms. On the other hand, restricting the refinement moveset to only pairing moves reduces SentRNA's performance to 53 / 63, slightly worse than ERD. The average GC content for each method is also reported in the rightmost column. For all other methods besides SentRNA, this value was reproduced from Garcia-Martin et al.[19] For SentRNA, the mean value was obtained by averaging over all valid solutions generated by SentRNA across all 154 models, and the reported error is the standard deviation.

| Puzzle | Pred | Ref | Time (s) | % Seq | Puzzle | Pred | Full | All pair | GC pair | Puzzle | % GC, Pred | % GC, Full | % GC, All pair | % GC, GC pair |
|---|---|---|---|---|---|---|---|---|---|---|---|---|---|---|
| RF00001.121 | | X | 0.595 | 0.923 | RF00001.121 | | X | X | X | RF00001.121 | | 0.5, 0.94 | 0.5, 0.76 | 0.91, 0.91 |
| RF00002.2 | | X | 3.337 | 0.987 | RF00002.2 | | X | X | X | RF00002.2 | | 0.58, 0.83 | 0.62, 1.0 | 0.67, 1.0 |
| RF00003.94 | | X | 6.289 | 0.95 | RF00003.94 | | X | | | RF00003.94 | | 0.57, 0.8 | | |
| RF00004.126 | X | X | 0.0 | 1.0 | RF00004.126 | X | X | X | X | RF00004.126 | 0.25, 0.89 | 0.2, 1.0 | 0.18, 1.0 | 0.23, 1.0 |
| RF00005.1 | | X | 0.459 | 0.973 | RF00005.1 | | X | X | X | RF00005.1 | | 0.29, 0.95 | 0.24, 1.0 | 0.33, 1.0 |
| RF00006.1 | X | X | 0.0 | 1.0 | RF00006.1 | X | X | X | X | RF00006.1 | 0.45, 1.0 | 0.3, 1.0 | 0.3, 1.0 | 0.4, 1.0 |
| RF00007.20 | X | X | 0.0 | 1.0 | RF00007.20 | X | X | X | X | RF00007.20 | 0.43, 0.76 | 0.31, 1.0 | 0.36, 1.0 | 0.33, 1.0 |
| RF00008.11 | X | X | 0.0 | 1.0 | RF00008.11 | X | X | X | X | RF00008.11 | 0.2, 1.0 | 0.2, 1.0 | 0.2, 1.0 | 0.2, 1.0 |
| RF00009.115 | | X | 5.412 | 0.983 | RF00009.115 | | X | X | X | RF00009.115 | | 0.4, 1.0 | 0.4, 0.98 | 0.47, 1.0 |
| RF00010.253 | | | inf | 0.0 | RF00010.253 | | | | | RF00010.253 | | | | |
| RF00011.18 | | | inf | 0.0 | RF00011.18 | | | | | RF00011.18 | | | | |
| RF00012.15 | X | X | 0.0 | 1.0 | RF00012.15 | X | X | X | X | RF00012.15 | 0.38, 1.0 | 0.26, 1.0 | 0.28, 1.0 | 0.33, 1.0 |
| RF00013.139 | X | X | 0.0 | 1.0 | RF00013.139 | X | X | X | X | RF00013.139 | 0.25, 1.0 | 0.25, 1.0 | 0.25, 1.0 | 0.25, 1.0 |
| RF00014.2 | X | X | 0.0 | 1.0 | RF00014.2 | X | X | X | X | RF00014.2 | 0.31, 1.0 | 0.24, 1.0 | 0.28, 1.0 | 0.31, 1.0 |
| RF00015.101 | X | X | 0.0 | 1.0 | RF00015.101 | X | X | X | X | RF00015.101 | 0.58, 1.0 | 0.42, 1.0 | 0.42, 1.0 | 0.48, 1.0 |
| RF00016.15 | | | inf | 0.0 | RF00016.15 | | | | | RF00016.15 | | | | |
| RF00017.90 | X | X | 0.0 | 1.0 | RF00017.90 | X | X | X | X | RF00017.90 | 0.44, 1.0 | 0.33, 1.0 | 0.34, 1.0 | 0.36, 1.0 |
| RF00018.2 | X | X | 0.0 | 1.0 | RF00018.2 | X | X | X | X | RF00018.2 | 0.49, 0.49 | 0.35, 0.96 | 0.37, 0.96 | 0.41, 1.0 |
| RF00019.115 | X | X | 0.0 | 1.0 | RF00019.115 | X | X | X | X | RF00019.115 | 0.39, 1.0 | 0.3, 1.0 | 0.39, 1.0 | 0.39, 1.0 |
| RF00020.107 | | | inf | 0.0 | RF00020.107 | | | | | RF00020.107 | | | | |
| RF00021.10 | X | X | 0.0 | 1.0 | RF00021.10 | X | X | X | X | RF00021.10 | 0.19, 1.0 | 0.16, 1.0 | 0.19, 1.0 | 0.19, 1.0 |
| RF00022.1 | X | X | 0.0 | 1.0 | RF00022.1 | X | X | X | X | RF00022.1 | 0.24, 1.0 | 0.21, 1.0 | 0.24, 1.0 | 0.24, 1.0 |
| RF00024.16 | | | inf | 0.0 | RF00024.16 | | | | | RF00024.16 | | | | |
| RF00025.12 | X | X | 0.0 | 1.0 | RF00025.12 | X | X | X | X | RF00025.12 | 0.43, 1.0 | 0.32, 1.0 | 0.32, 1.0 | 0.38, 1.0 |
| RF00026.1 | X | X | 0.0 | 1.0 | RF00026.1 | X | X | X | X | RF00026.1 | 0.6, 1.0 | 0.4, 1.0 | 0.4, 1.0 | 0.6, 1.0 |
| RF00027.7 | X | X | 0.0 | 1.0 | RF00027.7 | X | X | X | X | RF00027.7 | 0.16, 1.0 | 0.13, 1.0 | 0.13, 1.0 | 0.16, 1.0 |
| RF00028.1 | X | X | 0.0 | 1.0 | RF00028.1 | X | X | X | X | RF00028.1 | 0.45, 0.49 | 0.32, 0.96 | 0.3, 0.91 | 0.45, 1.0 |
| RF00029.107 | X | X | 0.0 | 1.0 | RF00029.107 | X | X | X | X | RF00029.107 | 0.47, 1.0 | 0.32, 1.0 | 0.26, 1.0 | 0.47, 1.0 |
| RF00030.30 | | X | 0.308 | 0.991 | RF00030.30 | | X | X | X | RF00030.30 | | 0.3, 0.98 | 0.27, 0.94 | 0.34, 0.91 |
| AB015827 | X | X | 0.0 | 1.0 | AB015827 | X | X | X | X | AB015827 | 0.45, 1.0 | 0.37, 1.0 | 0.38, 1.0 | 0.38, 1.0 |
| AF029195 | X | X | 0.0 | 1.0 | AF029195 | X | X | X | X | AF029195 | 0.48, 0.48 | 0.38, 0.98 | 0.43, 0.98 | 0.43, 0.99 |
| AF056938 | X | X | 0.0 | 1.0 | AF056938 | X | X | X | X | AF056938 | 0.41, 0.71 | 0.36, 1.0 | 0.36, 1.0 | 0.37, 1.0 |
| AF096836 | X | X | 0.0 | 1.0 | AF096836 | X | X | X | X | AF096836 | 0.4, 0.9 | 0.33, 1.0 | 0.34, 0.99 | 0.35, 1.0 |
| AF106618 | X | X | 0.0 | 1.0 | AF106618 | X | X | X | X | AF106618 | 0.32, 1.0 | 0.32, 1.0 | 0.32, 1.0 | 0.32, 1.0 |
| AF107506 | X | X | 0.0 | 1.0 | AF107506 | X | X | X | X | AF107506 | 0.5, 0.69 | 0.37, 0.99 | 0.37, 0.96 | 0.38, 1.0 |
| AF141485 | X | X | 0.0 | 1.0 | AF141485 | X | X | X | X | AF141485 | 0.46, 0.85 | 0.36, 0.99 | 0.39, 0.99 | 0.43, 1.0 |
| AJ011149 | | X | 2.763 | 0.981 | AJ011149 | | X | X | X | AJ011149 | | 0.3, 0.95 | 0.32, 0.94 | 0.33, 1.0 |
| AJ130779 | X | X | 0.0 | 1.0 | AJ130779 | X | X | X | X | AJ130779 | 0.55, 0.58 | 0.37, 0.99 | 0.42, 1.0 | 0.47, 1.0 |
| AJ132572 | X | X | 0.0 | 1.0 | AJ132572 | X | X | X | X | AJ132572 | 0.43, 0.64 | 0.34, 1.0 | 0.34, 1.0 | 0.43, 1.0 |
| AJ133622 | X | X | 0.0 | 1.0 | AJ133622 | X | X | X | X | AJ133622 | 0.49, 0.65 | 0.35, 0.68 | 0.34, 0.68 | 0.36, 0.72 |
| AJ236455 | | X | 26.464 | 0.977 | AJ236455 | | X | | | AJ236455 | | 0.34, 0.99 | | |
| D38777 | X | X | 0.0 | 1.0 | D38777 | X | X | X | X | D38777 | 0.76, 1.0 | 0.4, 1.0 | 0.36, 1.0 | 0.45, 1.0 |
| L11935 | X | X | 0.0 | 1.0 | L11935 | X | X | X | X | L11935 | 0.37, 1.0 | 0.33, 1.0 | 0.31, 1.0 | 0.36, 1.0 |
| LIU92530 | X | X | 0.0 | 1.0 | LIU92530 | X | X | X | X | LIU92530 | 0.5, 0.71 | 0.4, 0.84 | 0.48, 0.71 | 0.5, 0.73 |
| S70838 | X | X | 0.0 | 1.0 | S70838 | X | X | X | X | S70838 | 0.43, 1.0 | 0.3, 1.0 | 0.31, 1.0 | 0.34, 1.0 |
| U63350 | X | X | 0.0 | 1.0 | U63350 | X | X | X | X | U63350 | 0.35, 1.0 | 0.29, 1.0 | 0.3, 1.0 | 0.35, 1.0 |
| U81771 | X | X | 0.0 | 1.0 | U81771 | X | X | X | X | U81771 | 0.39, 1.0 | 0.35, 1.0 | 0.37, 1.0 | 0.39, 1.0 |
| U84629 | X | X | 0.0 | 1.0 | U84629 | X | X | X | X | U84629 | 0.49, 1.0 | 0.37, 1.0 | 0.38, 1.0 | 0.47, 1.0 |
| X61771 | | X | 128.68 | 0.941 | X61771 | | X | | | X61771 | | 0.39, 0.6 | | |
| X81949 | | X | 3.157 | 0.998 | X81949 | | X | X | X | X81949 | | 0.36, 1.0 | 0.32, 0.51 | 0.44, 0.9 |
| X99676 | X | X | 0.0 | 1.0 | X99676 | X | X | X | X | X99676 | 0.45, 0.5 | 0.42, 0.98 | 0.45, 0.98 | 0.45, 0.99 |
| Z83250 | X | X | 0.0 | 1.0 | Z83250 | X | X | X | X | Z83250 | 0.47, 0.73 | 0.34, 1.0 | 0.33, 0.97 | 0.33, 1.0 |
| 4 | X | X | 0.0 | 1.0 | 4 | X | X | X | X | 4 | 0.38, 1.0 | 0.2, 1.0 | 0.25, 1.0 | 0.25, 1.0 |
| 1 | X | X | 0.0 | 1.0 | 1 | X | X | X | X | 1 | 0.39, 1.0 | 0.28, 1.0 | 0.28, 1.0 | 0.39, 1.0 |
| 8 | X | X | 0.0 | 1.0 | 8 | X | X | X | X | 8 | 0.38, 0.38 | 0.29, 1.0 | 0.29, 0.99 | 0.31, 1.0 |
| 6 | X | X | 0.0 | 1.0 | 6 | X | X | X | X | 6 | 0.86, 0.86 | 0.29, 1.0 | 0.35, 0.99 | 0.33, 1.0 |
| 7 | X | X | 0.0 | 1.0 | 7 | X | X | X | X | 7 | 0.52, 0.84 | 0.31, 0.97 | 0.31, 0.98 | 0.38, 1.0 |
| 10 | X | X | 0.0 | 1.0 | 10 | X | X | X | X | 10 | 0.63, 0.63 | 0.28, 0.99 | 0.28, 0.99 | 0.3, 1.0 |
| 3 | X | X | 0.0 | 1.0 | 3 | X | X | X | X | 3 | 0.35, 1.0 | 0.29, 1.0 | 0.29, 1.0 | 0.32, 1.0 |
| 2 | X | X | 0.0 | 1.0 | 2 | X | X | X | X | 2 | 0.21, 1.0 | 0.12, 1.0 | 0.21, 1.0 | 0.21, 1.0 |
| L77117 | X | X | 0.0 | 1.0 | L77117 | X | X | X | X | L77117 | 0.71, 0.71 | 0.36, 0.71 | 0.47, 0.99 | 0.33, 1.0 |
| 5 | | X | 82.3 | 0.855 | 5 | | X | | | 5 | | 0.46, 0.7 | | |
| 9 | | | inf | 0.0 | 9 | | | | | 9 | | | | |

**Table 7:**
*Left:* Performance of the SentRNA ensemble of 154 trained models on the 63-structure test set of Garcia-Martin et al. while employing neural network prediction and the full refinement moveset (left). An "X" in the "Pred" column means that a model could predict a correct solution to the puzzle without refinement, and an "X" in the "Ref" column means the model was able to solve the puzzle after refinement. The number in the "Times" column represents the minimum refinement time across all models that was necessary to solve the puzzle, in seconds, and "% Seq" corresponds to the fraction sequence identity between the neural network prediction and the refined solution for that model. Refinement parameters were set to 300 adaptive walk trajectories of length 30, and results from refinement that exceeded 10 minutes were discarded. *Middle:* Comparison between the full refinement moveset and restricted movesets of GC and all pairing mutations. We observe that using the full refinement moveset allows SentRNA to solve strictly more targets (57 / 63) compared to using either GC pairing or the full set of pairing moves (53 / 63).

*Right:* A list of the range of GC content for each target across all models given a particular NN + refinement pipeline. "%GC, Pred" is the GC content from solutions generated using only the NN, "%GC, Full" is from NN + full refinement moveset, "%GC, All pair" is from NN + pairing moves, and "%GC, GC pair" is from NN + GC pairing moves. The minimum and maximum % GC content across all valid generated sequences for the target are reported. Depending on the model, the GC content for a sequence can vary dramatically, for example, ranging from 20% to 100% for the target RF00008.11.

Finally, we note that using a combination of neural network prediction and the full refinement moveset, SentRNA can solve one additional target that was unsolvable by all previously benchmarked algorithms given the 10-minute time limit (Figure 9).

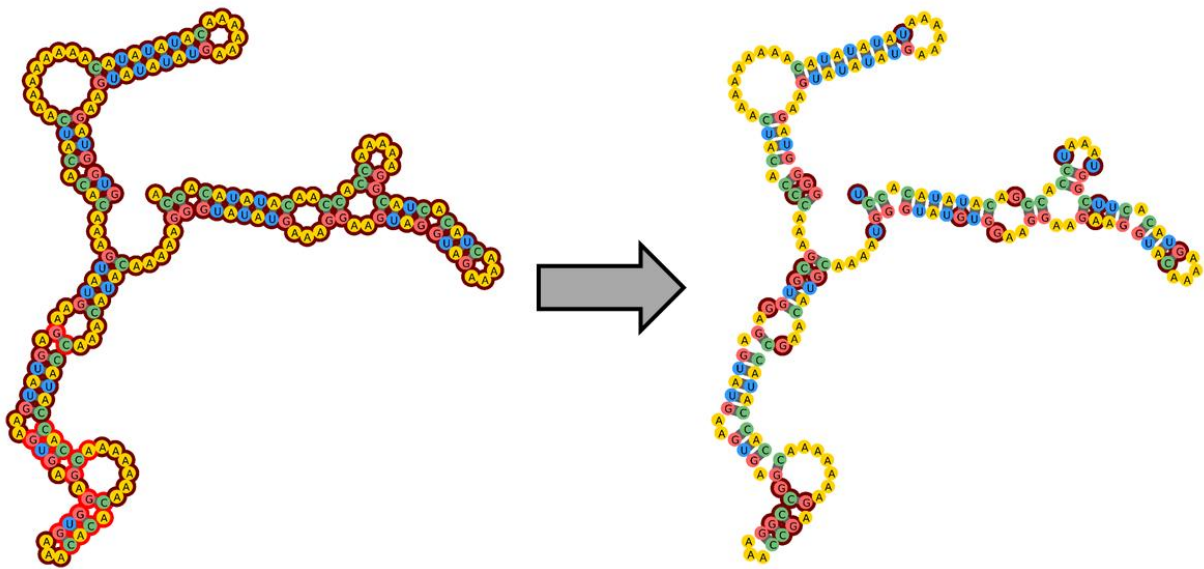

**Figure 9:** SentRNA can solve one target previously unsolvable by any computational algorithm within a 10-minute time limit through a combination of neural network prediction and refinement using the full moveset. The key stabilization strategies added by the refinement are boosting moves at the 2-3 loops using G-mutations.

We observe that the key stabilizations added by the refinement are a series of boosts to the small 2-3 internal loops using G mutations. SentRNA was unable to apply this strategy via the neural network prediction, indicating that it did not learn this strategy during training. Thus, once again we have an example in which the prior information incorporated in the refinement moveset can compensate for gaps in knowledge in SentRNA's neural network.

*Discussion:*
Our results show that incorporating a prior of human design strategies into an inverse RNA design agent can significantly boost its performance on difficult targets. We observe SentRNA's neural network's ability to incorporate advanced human strategies into the sequence initialization directly, such as the UGUG superboost for 2-2 loops, allows it to either predict a valid sequence immediately (e.g. *Mat – Cuboid*), or predict a sequence that is close to a valid solution, which can then be refined into a correct solution using only a basic refinement strategy that involves only pairing moves and encodes no additional human prior information (e.g. *Mat – Lot 2-2 B*). Interestingly, we observe that for several puzzles in the Eterna100, using the full refinement moveset to refine a neural network prediction is in fact not the optimal strategy, as this can lead to either irreversible damage of an intelligently initialized sequence (*Mat – Lot 2-2 B*), or unnecessary expansion of the refinement search space such that productive moves are sampled less often (*Bullseye*). In these situations, a simpler and more naïve refinement moveset can show significantly better performance. Overall, we observe that even using this much simpler refinement moveset (equivalent to the adaptive walk from RNAInverse) SentRNA can maintain a sizable advantage over other design algorithms, demonstrating that the neural network initialization of the sequence is indeed superior to the sequence initializations of other methods when solving the Eterna100.

Alternatively, for several puzzles SentRNA shows clear deficiencies in its neural network prediction, being unable to learn strategies such as zigzag stabilization to solve *hard Y*, or the single U-U boosting of thet 4-4 internal loop to solve *Misfolded Aptamer 6*. In these cases, the human prior information encoded in the unpairing and boosting moves of the refinement algorithm is often able to compensate for these slight deficiencies and solve the puzzle through brute-force search. However, we believe that there is no reason SentRNA cannot learn these strategies directly through training of its neural network, and hypothesize the reason SentRNA did not learn them during this study is because of their limited representation in the training set. For instance, U-U boosting is a more advanced form of boosting compared to boosting with G-mutations that is utilized by far fewer players overall, as well as being unnecessary to solve a large majority of the training set puzzles. Thus, due to the poor representation of U-U boosting when considering all $1.8 \times 10^4$ player solutions, it is no surprise that SentRNA has difficulty learning this strategy. However, if we used instead a more restricted training set consisting of only solutions from only the very best EteRNA players, who are more likely to have knowledge of and routinely use this advanced strategy, we believe SentRNA would be fully capable of learning it. Puzzles such as *Misfolded Aptamer 6* may then potentially be solvable using only pairing moves during refinement.

Overall, we notice that the success of SentRNA as an algorithm depends on synergistic behavior between the neural network prediction and the refinement process. When a large amount of human prior information is encoded into the neural network prediction, a simpler pairing-move-only refinement moveset that does not disrupt this prior information can lead to better performance. On the other hand, when the neural network prediction is more naïve, a more complex refinement moveset is necessary to compensate for this lack of knowledge. We stress that the combination of both elements is critical to SentRNA's success. Although the neural network may be able to solve simpler puzzles without any further refinement, more complicated puzzles (e.g. *Mat – Lot 2-2 B*) often contain structural nuances specific to that puzzle that are not represented in the training data. Therefore, some sort of refinement process is still necessary to sample a valid solution, even from a "close" sequence initialization.

Remarkably, we find that the strategies learned during training on EteRNA puzzles are useful for solving even a second, completely independent test set. When benchmarking on the 63 structures from Garcia-Martin et al.[19] we observed the neural network prediction could solve

46 / 63 targets, which already places in 3rd out of the 10 previously benchmarked methods. The strong performance of the neural network on its own given a completely independent test set suggests that the human-like strategies learned during training are indeed generalizable to unseen targets, and that we are not simply overfitting to EteRNA puzzles. Furthermore, if we include a naïve refinement strategy of pairing moves, we find SentRNA can solve 53 / 63, making it competitive with the current state-of-the-art, ERD, which solved 54 / 63. Finally, including the full refinement moveset allows SentRNA to gain a slight edge over ERD, solving 57 / 63. The fact that ERD performs better than SentRNA using the more naïve refinement suggests that perhaps ERD's initialization, which is drawn from a databse of known, naturally occurring RNA subsequences, is superior to SentRNA's "human" initialization for this test set. Alternatively, it could also be that because the evolutionary algorithm from ERD is more sophisticated than the simple adaptive walk employed by SentRNA, it is able to compensate for a potentially inferior initialization. Given these possibilities, creating a hybrid method such as SentRNA neural network + ERD evolutionary algorithm and seeing if it can surpass either individual method would be an interesting follow-up study.

Alternatively, adding additional moves to SentRNA's refinement moveset, such as explicit moves for zigzag stabilization, could also potentially improve SentRNA's current performance by allowing it to solve puzzles such as *hard Y* more quickly. However, this comes with the cost of adding more ways to potentially disrupt an intelligent neural network sequence initialization, as was the case with puzzles such as *Mat – Lot 2-2 B*. Given our results, we hypothesize that instead of expanding the refinement moveset, the best way to improve SentRNA's performance is to feed it more instructive training data such that these advanced strategies can be encoded into the neural network prediction itself, and *simplify* the moveset as much as possible (i.e. to only pairing moves) to reduce the size of the refinement's search space.

Finally, we note that because the ultimate goal in solving inverse RNA folding is to assist real-world RNA design, training SentRNA using player solutions whose folds have been judged not only using the *in silico* ViennaRNA energy function, but also which have been experimentally validated, might allow our agent to more effectively learn strategies for real-world RNA design. In fact, it has already been shown that incorporating human design strategies into a computational agent can allow it to achieve state-of-the-art performance in this regard. Previously, EteRNA developers created a design agent called EteRNABot, which combined a set of 40 player-submitted, experimentally validated rules and strategies into a custom score function, which EteRNABot then tried to optimize using a stochastic search algorithm similar to that of RNAInverse. Through this approach, EteRNABot could globally outperform RNAInverse and NUPACK on a set of nine difficult design challenges, judged based on consistency with actual experimental synthesis and structure mapping studies.[24] We posit that SentRNA, if trained on experimentally validated player solutions, could yield comparable, if not better, results since it learns in a purely data driven manner and can therefore potentially learn a much richer set of strategies from the data instead of relying on a limited set of hand-coded strategies.

**Conclusion:**
We present SentRNA, a computational agent for RNA design that consists of a fully-connected neural network trained on player-submitted solutions from the online RNA design game EteRNA, coupled with an adaptive walk algorithm that incorporates simple human design strategies.

Given a target structure, the agent predicts a sequence that folds into that structure, which is then further refined via the adaptive walk if necessary. We observe that SentRNA can effectively learn and apply human-like design strategies to achieve state-of-the-art performance on the difficult Eterna100 test set, solving 78 / 100 puzzles in total. Furthermore, we show that the strategies SentRNA learns are generalizable to non-EteRNA targets, achieving state-of-the-art performance on an independent test set of 63 non-EteRNA targets, solving 57 / 63 targets. Our results demonstrate the power of incorporating human prior information into a design algorithm, and suggests a new paradigm in machine-based RNA design.

**Acknowledgements:**
We would like to thank the EteRNA community for providing the training data for SentRNA and making this study possible. We would also like to thank Tono Garcia-Martin and Prof. Peter Clote for sharing their compiled set of 63 non-EteRNA targets as well as their benchmarking data for these targets.

**Author contributions:**
J.S. programmed SentRNA, trained models and evaluated their performance, and wrote the paper. EteRNA players generated the *eternasolves* dataset used to train SentRNA. R. D. and V. S. P. supervised the project.

**References:**

[1] Goldberg, MS, Xing, D, Ren, Y, Orsulic, S, Bhatia, SN, Sharp, PA. Nanoparticle-mediated delivery of siRNA targeting Parp1 extends survival of mice bearing tumors derived from Brca1-deficient ovarian cancer cells. PNAS. 2010:108:745-750.

[2] Win, MN, Smolke, CD. Higher-Order Cellular Information Processing with Synthetic RNA Devices. Science. 2008:322:456-460.

[3] Delebecque, CJ, Lindner, AB, Silver, PA, Aldaye, FA. Organization of Intracellular Reactions with Rationally Designed RNA Assemblies. Science. 2011:333:470-473.

[4] Hao, C, Li, X, Tian, C, Jiang, W, Wang, G, Mao, C. Construction of RNA nanocages by re-engineering the packaging RNA of Phi29 bacteriophage. Nature Communications. 2014:5:1-7.

[5] Dixon, N, Duncan, JN, Geerlings, T, Dunstan, MS, McCarthy, JEG, Leys, D et al. Reengineering orthogonally selective riboswitches. PNAS. 2010:107:2830-2835.

[6] Hofacker IL, Fontana W, Stadler PF, Bonhoeffer LS, Tacker M, Schuster P. Fast folding and comparison of RNA secondary structures. Monatshefte Für Chem Chem Mon. 1994;125: 167–188. doi:10.1007/BF00818163

[7] Andronescu M, Fejes AP, Hutter F, Hoos HH, Condon A. A New Algorithm for RNA Secondary Structure Design. J Mol Biol. 2004;336: 607–624. doi:10.1016/j.jmb.2003.12.041

[8] Busch A, Backofen R. INFO-RNA--a fast approach to inverse RNA folding. Bioinformatics. 2006;22: 1823–1831. doi:10.1093/bioinformatics/btl194

[9] Matthies MC, Bienert S, Torda AE. Dynamics in Sequence Space for RNA Secondary Structure Design. J Chem Theory Comput. 2012;8: 3663–3670. doi:10.1021/ct300267j

[10] Zadeh JN, Steenberg CD, Bois JS, Wolfe BR, Pierce MB, Khan AR, et al. NUPACK: Analysis and design of nucleic acid systems. J Comput Chem. 2011;32: 170–173. doi:10.1002/jcc.21596

[11] Taneda A. MODENA: a multi-objective RNA inverse folding. Adv Appl Bioinforma Chem AABC. 2010;4: 1–12.

[12] Reinharz, V et al. A weighted sampling algorithm for the design of RNA sequences with targeted secondary structure and nucleotide distribution. Bioinformatics. 2013;29:308-315.

[13] Lyngs, R. B. et al. Frnakenstein: multiple target inverse RNA folding. BMC Bioinformatics. 2012;13:260.

[14] Esmaili-Taheri, A. and Ganjtabesh, M. ERD: A fast and reliable tool for RNA design including constraints. BMC Bioinformatics, 2015;16:20.

[15] Garcia-Martin, JA, Clote, P, and Dotu, I. RNAiFOLD: a constraint programming algorithm for RNA inverse folding and molecular design. J Bioinform Comput Biol. 2013;11: 1350001.

[16] Kleinkauf, R, Mann, M, Backofen, R. antaRNA: ant colony-based RNA sequence design. Bioinformatics. 2015;31:3114-3121.

[17] Lee, JA, Fisker, E, Kosaraju, V, Wu, M, Kong, J, Lee, J et al. Principles for Predicting RNA Secondary Structure Design Difficulty. J. Mol. Biol. 2016:428:748-757.

[18] EteRNA is hosted at www.eternagame.org

[19] Garcia-Martin, JA et al. RNAiFold 2.0: a web server and software to design custom and Rfam-based RNA molecules. Nucleic Acids Research. 2015;43:513-521.

[20] Sherlock 2.0 documentation can be found at: https://www.sherlock.stanford.edu/docs/overview/transition

[21] Abadi, M, Agarwal, A, Barham, P, Brevdo, E, Chen, Z, Citro, C et al. TensorFlow: Large-scale machine learning on heterogeneous systems, 2015. Software available at tensorflow.org

[22] Kingma, D. P. and Ba, Jimmy. Adam: A Method of Stochastic Optimization, arXiv:1412:6980, 2014.

[23] Beginner boosting tutorial: http://eternagame.wikia.com/wiki/Boosting

[24] Lee, JA, Kladwang, W, Lee, M, Cantu, B, Azizyan, M, Kim, H et al. RNA design rules from a massive open laboratory. Proc. Nat. Acad. Sci. 2013:112:2122-2127.